%% file: main.tex
\documentclass[aps,pra,twocolumn,superscriptaddress,longbibliography]{revtex4-1} 
\usepackage{subfigure}
\usepackage{hyperref}
\usepackage[pdftex]{color}
\usepackage{soul}
\input{header}

\hypersetup{
    colorlinks=true,       
    linkcolor=cyan,          
    citecolor=magenta,        
    filecolor=magenta,      
    urlcolor=cyan,           
    runcolor=cyan
}

\begin{document}
\title{Ferromagnetically shifting the power of pausing}
\author{Zoe Gonzalez Izquierdo}
\affiliation{Department of Physics and Astronomy, and Center for Quantum Information Science \& Technology, University of Southern California, Los Angeles, California 90089, USA}
\affiliation{QuAIL, NASA Ames Research Center, Moffett Field, California 94035, USA}
\affiliation{USRA Research Institute for Advanced Computer Science, Mountain View, California 94043, USA}

\author{Shon Grabbe}
\affiliation{QuAIL, NASA Ames Research Center, Moffett Field, California 94035, USA}
\author{Stuart Hadfield}
\affiliation{QuAIL, NASA Ames Research Center, Moffett Field, California 94035, USA}
\affiliation{USRA Research Institute for Advanced Computer Science, Mountain View, California 94043, USA}
\author{Jeffrey Marshall}
\affiliation{QuAIL, NASA Ames Research Center, Moffett Field, California 94035, USA}
\affiliation{USRA Research Institute for Advanced Computer Science, Mountain View, California 94043, USA}
\author{Zhihui Wang}
\affiliation{QuAIL, NASA Ames Research Center, Moffett Field, California 94035, USA}
\affiliation{USRA Research Institute for Advanced Computer Science, Mountain View, California 94043, USA}
\author{Eleanor Rieffel}
\affiliation{QuAIL, NASA Ames Research Center, Moffett Field, California 94035, USA}
\date{\today} 

\begin{abstract}
We study the interplay between quantum annealing parameters in embedded problems, providing both deeper insights into the physics of these devices and pragmatic recommendations to improve performance on optimization  
problems. We choose as our test case 
the class of degree-bounded minimum spanning tree problems.  
Through runs on a D-Wave quantum annealer, we demonstrate that pausing in a specific time window in the anneal provides 
improvement in the probability of success and in the time-to-solution for these problems.
The time window is consistent across problem instances, and its location is within the region suggested by prior theory and seen in 
previous results on native problems. An approach to enable gauge transformations for problems with the qubit coupling strength $J$ in an asymmetric range is presented and shown to significantly improve performance.
We also confirm that the optimal pause location exhibits a shift with the magnitude of the ferromagnetic coupling, $|J_F|$, between physical qubits representing the same logical one.
We extend the theoretical picture for pausing and thermalization in quantum annealing to the embedded case. This picture, along with perturbation theory analysis, and exact numerical results on small problems, confirms that the effective pause region moves earlier in the anneal as $|J_F|$ increases.
It also suggests why pausing, while still providing significant benefit, has a less pronounced effect on embedded problems. 

\end{abstract}

\maketitle
\section{Introduction}

Quantum computing provides novel mechanisms for efficient computing, but the extent of its impact is as of yet undetermined. A tantalizing area of application is combinatorial optimization, where challenging instances are currently attacked by a variety of classical heuristics, and where quantum heuristics have the potential to outperform these classical approaches. Here, we advance the understanding of one such heuristic, quantum annealing, deepening the theoretical picture of the roles that thermalization, adiabatic processes, and diabatic process play in quantum annealing, and demonstrating the impact of annealing schedules and the interplay between quantum annealing parameters on performance, particularly on application-related problems that require embedding.

Our work builds on the theoretical picture of Marshall et al.~\cite{Marshall19_Pausing} that explains why pausing in an appropriate time window during the anneal enables the system to thermalize better, improving the fit of the output distribution with a Boltzmann distribution and increasing the success probability by orders of magnitude. 
Because quantum annealing happens at non-zero temperature, temperature plays a significant role, along with quantum dynamics induced by varying the Hamiltonian, particularly near where the temperature and the minimal energy gap between the ground state and the first excited state are commensurate. 
This effect has also been studied in simulations \cite{p-spin-pause}, and recently, rigorous sufficient conditions under which pausing helps were identified in Ref.~\cite{chen-pausing}.
Here, we build on the above understanding, beyond the native problems studied in \cite{Marshall19_Pausing}, to embedded problems. 

It is well known that most problem instances, in particular those related to applications, will not have a structure that matches that of the hardware, in which case the problems must be embedded. Embedded problems use multiple physical qubits to represent each logical qubit, with these physical qubits coupled via ferromagnetic couplings $J_F<0$. In the embedded problems we study, we confirm an improvement in success probability and that for this class of problem, as was found for native problems, there is a time window in which a pause reliably improves the performance across problem instances. We extend the theoretical picture of Refs.~\cite{Marshall19_Pausing,chen-pausing} to embedded problems, including a perturbative analysis on the effect of $|J_F|$ on the minimal energy gap between the ground and the first excited states. Our gap analysis and numerical simulations on small systems show that as $|J_F|$ increases, the minimal gap shifts earlier and the gap size decreases. This extended picture explains why one would expect a shift of the optimal pause location to earlier in the anneal with increasing $|J_F|$, and also a somewhat less pronounced improvement from pausing on embedded problems than on native problems.

The class of problems studied, bounded-degree minimal spanning tree (BD-MST) problems, seen in a variety of application areas such as a broad spectrum of network-related problems, have not been studied before in this context. We demonstrate that small instances of these problems can be embedded and successfully solved by state-of-the-art quantum annealers, and confirm the results predicted by our theoretical picture. We demonstrate that for the best parameters, pausing improves not only the probability of success, but also the time-to-solution (TTS). To obtain these results, we used the newly added extended range feature of the D-Wave 2000Q to enable the use of stronger ferromagnetic couplings relative to the problem instance couplings. Because of the asymmetry in the extended range, we could not use the standard gauge approach to randomize the effect of qubit biases in the D-Wave 2000Q on the annealing runs. We developed a partial gauge approach that enabled us to obtain much cleaner results and substantially better probabilities of success than running without partial gauges.

The rest of the paper is organized as follows:
In Sec.~\ref{sec:background}, we review background information on spanning tree problems and on quantum annealing.
In Sec.~\ref{sec:methods}, we describe the specifics related to the hardware, the instances, and the parameters for our runs, and the metrics we use to evaluate them.
Sec.~\ref{sec:results} is devoted to results on the annealer.
Results for annealing without pause are shown in Sec.~\ref{sec:base_case},
how pausing can be helpful is demonstrated in Sec.~\ref{sec:improve_with_pause}
and how pausing shifts with $|J_F|$ in Sec.~\ref{sec:shift}.
The technical treatment that enables the conclusive results, partial gauges, is discussed in Sec.~\ref{sec:partial_gauges}.
We provide theoretical analysis and a physical picture for the shifting of optimal pause location with $|J_F|$ in Sec.~\ref{sec:pictures}.  In Sec.~\ref{sec:conclusion} we summarize the results and discuss future work.

\section{Background}
\label{sec:background}

We review background material on spanning tree problems and on quantum annealing.

\subsection{Spanning Tree Problem Classes} 

\begin{definition}
A spanning tree for a graph $G$ is a subgraph of $G$ that is a tree and contains all vertices of~$G$.
\end{definition}

Spanning trees are important for several reasons.  They play a critical role in designing efficient routing algorithms.  
Some computationally hard problems, such as the Steiner Tree problem and the Traveling Salesperson Problem, can be solved approximately
using spanning trees~\cite{vazirani2013approximation}. 
Spanning tree problems also find broad applications in network design, bioinformatics, etc. 

One flavor of the spanning tree problems is the weighted spanning tree problem:
Given a connected undirected graph $G=(V,E)$ and set of weights $w_{uv}$ for each edge $(uv)\in E$, we seek a spanning tree $T\subset E$ such that the tree weight $\sum_{(uv) \in T} w_{uv}$ is minimized. 

For general graphs, determining if there exists a spanning tree of weight $W$ can be decided in polynomial time, and different efficient algorithms exist to find a minimum weight tree; 
for example, Kruskal's algorithm requires time $O(|E| \log |V|)$  \cite{cormen2009introduction}. (Special classes of graphs can be solved even faster.) 
On the other hand, with the additional constraint that the maximum vertex degree of the spanning tree found is at most $\Delta$, even deciding whether there exists such a spanning tree becomes NP-complete for fixed $\Delta\ge 2$ \cite{GareyJohnson}. 
In this work we focus on the bounded-degree maximum spanning tree (BD-MST) problem:

\textbf{The BD-MST Problem:}
Given an integer $\Delta\ge 2$ and graph $G=(V,E)$ with edge weights $w_{uv}$, $(uv)\in E$,  find a minimum weight spanning tree of maximum degree at most~$\Delta$.

We refer interested readers to Appendix.~\ref{sec:complexity} and References therein for approximation complexity
theory related to the BD-MST problem. 

\subsection{Solving on a Quantum Annealer}

Quantum annealing is a quantum metaheuristic for optimization. Quantum annealers are quantum hardware that are designed to run this metaheuristic. Any classical cost function $C(x)$ that is a polynomial over binary variables $x\in\{0,1\}^n$ can, with the addition of auxiliary variables, be turned into a quadratic cost function. Problems with quadratic cost functions over binary variable without additional constraints are called
quadratic unconstrained binary optimization 
(QUBO) problems. Quantum annealing is carried out by evolving the system under a time-dependent Hamiltonian
$H(s) = A(s)H_D + B(s)H_C$, 
where $H_D$ is a driver Hamiltonian, most commonly $H_X = -\sum_i X_i$ and $H_C$ is an Ising Hamiltonian derived from a classical cost function. There is a straightforward mapping between QUBO  and Ising problems. 
The parameter $s$ is a dimensionless time parameter that ranges from $0$ to $1$, with $A(s)$ and $B(s)$ determining the form of the anneal schedule. As we will see, many different schedules $s(t)$ are possible. More information about quantum annealing generally, including mappings of problems to QUBO can be found in \cite{Choi19, Rieffel14CaseStudy,STMapping}.

For most application problems, on a hardware with restricted qubit connectivity, the resulting QUBO problem must further be embedded to conform with the hardware connectivity;
graph minor embedding enables coupling between  logical qubits in the QUBO graph by representing each logical qubit by a set of physical qubits with ferromagnetically coupled with a magnitude of $|J_F|$ among them to promote collective behavior ($J_F$ is always negative, so we typically refer to its magnitude $|J_F|$.). Following standard terminology in graph theory, each such set of physical qubits is called a {\it vertex model} for its corresponding logical qubit.
When embedding, we use the same coupling strength $|J_F|$ for all the couplings within a vertex model. 
Problems that do not require embedding because their structure matches that of the hardware are called {\it native problems} for
that hardware.

While $|J_F|$ can be set to a large value such that the embedded problem preserves the ground state of the logical problem, 
and analytical bounds on this value can be obtained~\cite{Choi}, too large a $|J_F|$ can reduce quantum annealing performance. 
Physically there is an energy limit on the Hamiltonian as a whole, and too large a $|J_F|$ relative to other parameters would mean that all of the problem parameters could reduce performance due to precision issues and noise in implementation.
Furthermore, the energy spectrum throughout the anneal varies with the value of $|J_F|$, and its effect on the annealing often requires careful case-by-case consideration~\cite{Choi19, Rieffel14CaseStudy, Warburton, Venturelli15}.
Thus, optimally setting the ferromagnetic coupling $|J_F|$ 
is a challenging task. 
Prior work has shown there is a sweet spot for this value.
Physically this makes sense because a stronger $|J_F|$
makes it less likely for individual qubits within a vertex model to flip, which helps to avoid breaking the vertex model, but too large a $|J_F|$ makes it increasingly costly for the vertex model qubit values to flip together, potentially preventing the system from leaving a non-optimal configuration. 

To boost the probability of success, $|J_F|$ must strike the right balance, leading to better chances of arriving at---and staying in---the correct configuration.
The D-Wave 2000Q allows asymmetric extension of the pairwise qubit coupling strengths 
$J_{i,j} \in [-2,1]$  
(in addition to the canonical symmetric option $J_{i,j}\in[-1,1]$).  One usage of this extension is to set $|J_F|$ in the extended range.  We show how the extended values improve the success probability of our problems.

The schedule $s(t)$ can significantly affect performance. Of particular interest to us are schedules that include a pause where for some sub-interval $s(t)$ is constant
(i.e., $H$ is constant for a specified time). 
Marshall et al.~\cite{Marshall19_Pausing} observed on an ensemble of native problems that, strikingly, a pause at a location (generally) insensitive to the instance
specifics boosts the probability of finding the ground state---the success probability---by orders of magnitude. 
The physical picture underlying such a universal effect is reviewed and expanded in Sec.~\ref{sec:pictures}

\section{Methods}
\label{sec:methods}

Here, we discuss the specifics of the problem instances, annealing schedules and parameters, and metrics used to obtain our results.

\subsection{Problem Instances}

Each \mst problem instance consists of a weighted graph $G=(V,E)$ and a degree bound $\Delta$. The underlying graphs are chosen by exhausting all connected graphs with $n=|V|=5$, which have $m=|E|$ ranging from 4 to 10. Weight sets were uniformly drawn from $1$ to $7$.  Graphs and weight sets were combined to yield a large number of unique instances. Results are averaged over ensembles of instances. The size of the ensemble will be specified 
for each result in Section~\ref{sec:results}.
The complete list of graphs and weight sets can be found in Table~\ref{tab:n5graph} and Table~\ref{tab:weights} respectively of Appendix \ref{app:instances}. 

A number of mappings of the \mst problem to QUBO can be found in Ref.~\cite{wang2020}; here we use the resource-efficient level-based mapping
described in App.~\ref{appendix:mapping}.
For each problem instance, the level-based mapping yields an objective function Hamiltonian $H_C$.
For the degree bound we generally selected $\Delta=2$, resulting in problems equivalent to Hamiltonian path problems; we also tested
$\Delta=3$
and our claims hold for this case as well (see Fig.~\ref{fig:delta_3} in Appendix~\ref{appendix:more}).

\subsection{Annealing Parameters and Schedules}
\label{annealing_params}

We ran our problems on the D-Wave 2000Q quantum annealer housed at NASA Ames Research Center, which has $2031$ qubits and 
a Chimera graph architecture~\cite{DWave}.
To embed the resulting QUBO instances in the D-Wave 2000Q hardware graph, we ran D-Wave's
embedding-finding algorithm 30 times and used the smallest size embedding found (fewest total physical qubits). 
This procedure found an embedding for all graphs we considered.
Detailed information about the typical size of the embedded problems for different graphs can be found in Fig.~\ref{fig:chimera_vs_pegasus} in Appendix~\ref{appendix:embedding}, including the number of physical qubits and the size of the vertex models. Embedding statistics for a future D-Wave architecture (Pegasus) are also given.

The objective Hamiltonians were scaled so that the coupling strengths are in the range $[-1,1]$. 
In the embedded Hamiltonian, the extended $J$ range is used to couple physical qubits representing the same logical qubits. We chose a $|J_F|$ in the range $[1, 2]$, initially exploring all values in that range at $0.1$ intervals.

We used the D-Wave 2000Q default $A(s)$ and $B(s)$, exploring
two qualitatively different schedules. The first is a standard anneal, with time parameter 
$s(t) = t/t_a$, where $t_a$ is the annealing time. Baseline runs were performed with this schedule, and several annealing times $t_a$ initially tested. The shortest time allowed by D-Wave, $t_a = 1~\mu$s, was found to be optimal in terms of TTS for the instance ensembles, agreeing with previous studies for other problems~\cite{Boixo2014evidence, Ronnow2014defining, Venturelli15}.

The second type of schedule includes a pause. The beginning and end of the anneal are the same as in the first case, but at some intermediate point $s_p$ the Hamiltonian is held constant for some time $t_p$.
The entire range of possible pause locations ($s_p \in [0,1]$) was initially surveyed. A peak was reliably found
(see Sec.~\ref{sec:improve_with_pause}). Although the location of the peak is affected by $|J_F|$, it is always within the range $[0.2, 0.5]$, so further runs were limited to this region of interest, with $s_p$ varied between $0.2$ and $0.5$ at $0.02$ intervals.
A range of pause durations $t_p$ were also surveyed. Since shorter pause times found to yield better TTS 
(see Sec.~\ref{sec:improve_with_pause}), 
our runs are performed with pause duration $t_p = 1 \mu$s unless otherwise noted. After optimal values for other parameters were found, other $t_p$ values in the range $[0.25, 2] \mu$s were explored. 

We use the extended range of $J_{i,j}\in[-1,2]$ (in addition to the canonical symmetric option $J_{i,j}\in[-1,1]$). The asymmetry in the range w.r.t zero yields invalid a general strategy, gauge transformation (or, spin reversal transformation), which has been shown very effective in reducing noise effects and 
obtaining higher quality output data.  We designed and implemented a novel strategy, \emph{partial gauge transformation}, that selectively applies the transformation only to couplings in the symmetric range $[-1,1]$.  For the case that only the embedding couplings are in the extended range, this is equivalent to applying a general gauge transformation to the logical problem prior to the embedding, and is simple to implement. We found that 
the partial gauge transformation significantly helped in both boosting the success probability and reducing the output variance. 
Only by employing partial gauges could we obtain results clean enough to see various features we report on, such as
the positive role of an extended $|J_F|$ in the case of no pausing, and the shift of the optimal pause location with $|J_F|$. 
Partial gauges, and their effect, will be discussed in more detail in Sec.~\ref{sec:partial_gauges}.

Unless otherwise specified, all runs are performed with $t_a = 1~\mu$s, $50,000$ anneals (or reads), and 100 partial gauges.

\subsection{Metrics}
\label{sec:metrics}

We use the empirical probability of success ($p_{success}$) and time to solution (TTS) as our figures of merit for determining how likely a problem is to be solved, defined as:
\begin{align}
    & p_{success} = \frac{\text{\texttt{\#} anneals with correct solution}}{\text{total \texttt{\#} anneals}}
\\
& \mathrm{TTS} = \frac{\log(1 - 0.99)}{\log(1.0 -  p_{success})}t_{tot}\;,
\label{eq:TTS}
\end{align}
where the total time $t_{tot}=t_a+t_p$ is the time spent on each run, taking into account both the base annealing time $t_a$ and the pause duration $t_p$. 

These two measures are complementary to each other. The TTS figure of merit reports the expected time required to solve the problem with $99\%$ confidence.
While $p_{success}$ is directly determined by and hence provides a portal to understand the underlying physical process,  TTS gives a more practical measure that is universal across different parameter ranges and different solvers.
A higher success probability does not necessarily mean a higher TTS.  For instance, we might get a slightly higher $p_{success}$ by using a longer annealing time $t_a = 100 \mu$s than a shorter one $t_a = 1 \mu$s, 
yet the chance of finding the solution might be higher by repeating the $t_a = 1 \mu$s runs 100 times than doing the 
$t_a = 100 \mu$s anneal once.

Because we compare results from two different schedules (baseline no-pause and pause), we also need metrics that help us examine the benefits that the latter presents over the former. To this end, we define two quantities based on the instance-wise improvement in TTS.
The first one is the absolute TTS improvement, defined for each instance $i$ as
\begin{align}
\Delta \text{TTS}_i = \text{TTS}_i \text{(no pause) - TTS}_i \text{(pause)}\;, 
\end{align}
with the two TTS values calculated at their respective optimal $|J_F|$ values ($|J_F^*| = 1.6$ for the no pause case and $1.8$ for the pause case). A positive $\Delta \text{TTS}_i$ indicates that a pause improves upon the baseline results (i.e. reduces TTS) for that particular instance.
The second one is the relative TTS improvement, defined as the ratio
\begin{align}
\Delta \text{TTS}_i / \text{TTS}_i = \frac{\text{TTS}_i \text{(no pause) - TTS}_i \text{(pause)}}{\text{TTS}_i\text{(no pause)}}\;. 
\end{align}

When a valid solution is not found for a specific instance, and thus $p_{success}=0$ for that instance, its corresponding TTS is infinity. 
If TTS for both the pause and no pause results are infinity, the pause is not improving upon the no pause results, hence $\Delta \text{TTS}_i= 0$. When TTS$=\infty$ only for the no pause case, pausing provides the maximum possible improvement, and we set $\Delta \text{TTS}_i = \infty$ and $\Delta \text{TTS}_i / \text{TTS}_i = 1$. Finally, when TTS$=\infty$ only for the pause case, the opposite occurs, with $\Delta \text{TTS}_i = -\infty$ and $\Delta \text{TTS}_i / \text{TTS}_i = -\infty$.

After the embedded problem is run on the D-Wave, outputs with any inconsistent values on physical qubits that represent the same logical qubit--or with violated penalty terms such that the output doesn't encode a degree bounded spanning tree--are considered to be invalid answers, and counted as failed runs. The retained valid answers are then verified against the exact solution of the problem, which is obtained through direct enumeration for the 
small problem sizes we consider.
Reported data points correspond to the median, with the error bars marking the $35^{\mathrm{th}}$ and $65^{\mathrm{th}}$ percentiles. 
For ensembles of instances, $10^5$ bootstraps 
are performed over the instances to obtain those values, where each bootstrap sample is drawn with replacement from the original instance ensemble until it and is of the same size 
as the original ensemble.
Median and 35 and 65 percentiles from the bootstrap samples are reported.
There are a few instances that did not solve  with or without pauses; these instances are are not excluded from the ensemble in our bootstrap procedure, but are given a TTS of $\infty$.
These $\pm \infty$ values for $\Delta \text{TTS}_i$ and $\Delta \text{TTS}_i / \text{TTS}_i$ do not appear in our reported results, as they remain very far from the median (which we report as our data point) and from the $35^{th}$ and $65^{th}$ percentiles of the bootstrapped results that we present as error bars.

D-Wave returns the solution with the minimum cost it has found. To ensure the validity of this solution, we first confirm that the resulting graph is in fact a spanning tree that satisfies the degree constraint, and also a true optimal solution by comparing with the true minimal cost obtained by an exact classical algorithm.  Any other outcome is weighted zero toward $p_{success}$.

\section{Results}
\label{sec:results}
We now present our results on D-Wave 2000Q, including anneals without a pause (baseline) and the effect of pausing.

\subsection{Annealing without pause, effect of $|J_F|$}
\label{sec:base_case}

We first show that the \mst problems we study are successfully solved on the D-Wave 2000Q using a standard annealing schedule, demonstrating the ability of a quantum annealer to solve a new class of optimization problems,
and study the effect of the strength of the ferromagnetic coupling on the success probability.

\begin{figure}[h]
\includegraphics[width=\linewidth]{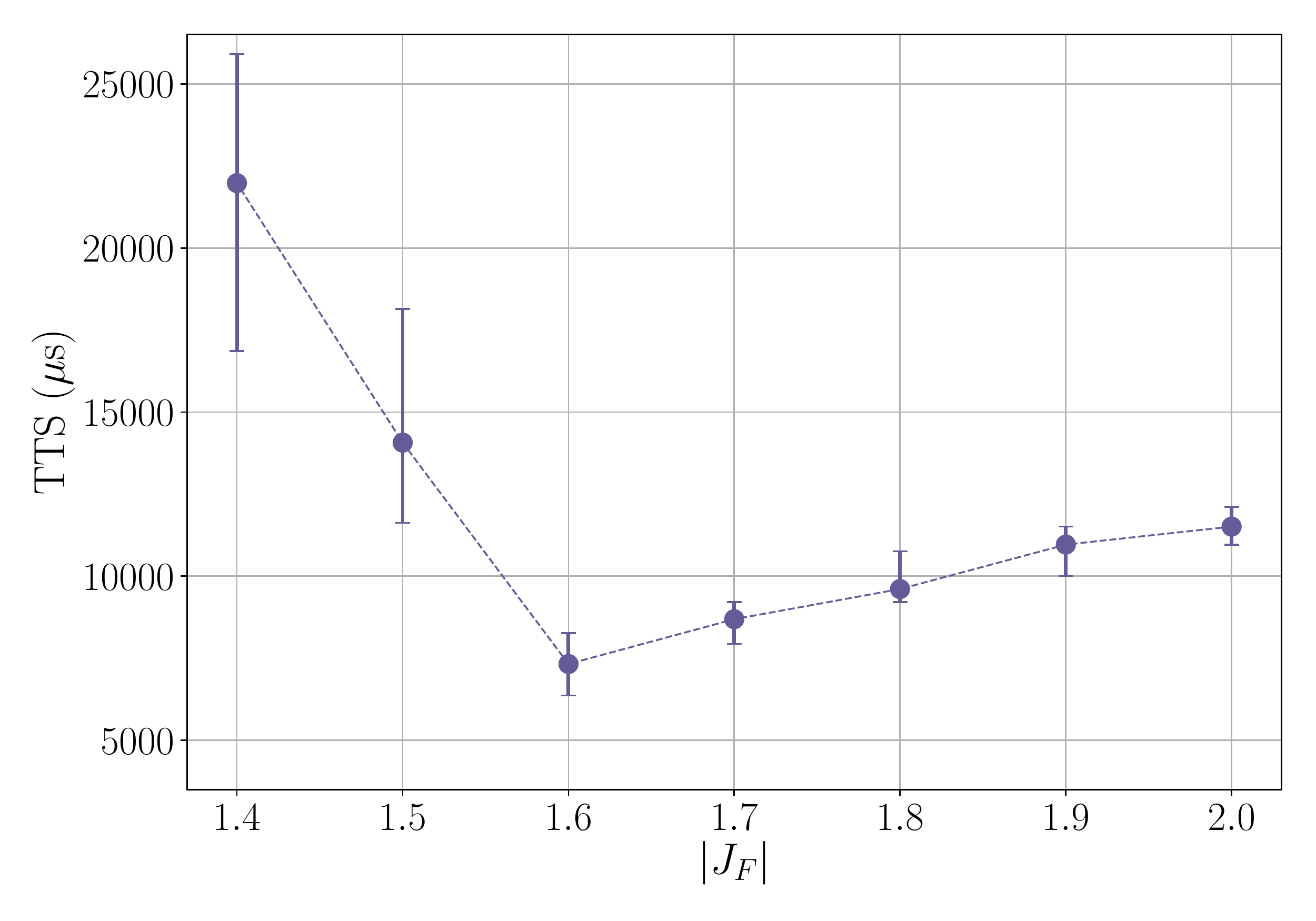}
  \caption{{\bf Optimal $|J_F|$ for baseline.} 
  TTS for an ensemble of 45 instances as $|J_F|$ varies.
  1 $\mu$s anneal is used.
  The best performance is observed at $|J_F^*|=1.6$
  }
  \label{fig:vary_Jf}
\end{figure}

The \emph{baseline} results are obtained with no pause and $t_a = 1 \mu s$, which is the shortest that D-Wave allows, and was chosen for consistently yielding the best TTS for ensembles of problem instances for both this study and previously studied problems.~\cite{Boixo2014evidence, Ronnow2014defining, Venturelli15}

By exploring the available range of $|J_F|$ values between $1.0$ and $2.0$, 
we confirm the advantage of using the extended $|J_F|$ range and identify its optimal value for the base case at $|J_F^*|=1.6$ with statistical significance, as shown in Fig.~\ref{fig:vary_Jf} where the success probability and TTS are shown for a range of $|J_F|$ for the ensemble of instances.

Results vary for groups of instances with different $n$; the optimal $|J_F|$ for $n=4$ is lower, around $1.2$ or $1.3$.

\subsection{Improvement with a pause} 
\label{sec:improve_with_pause}

\begin{figure}[h]
\includegraphics[width=\linewidth]{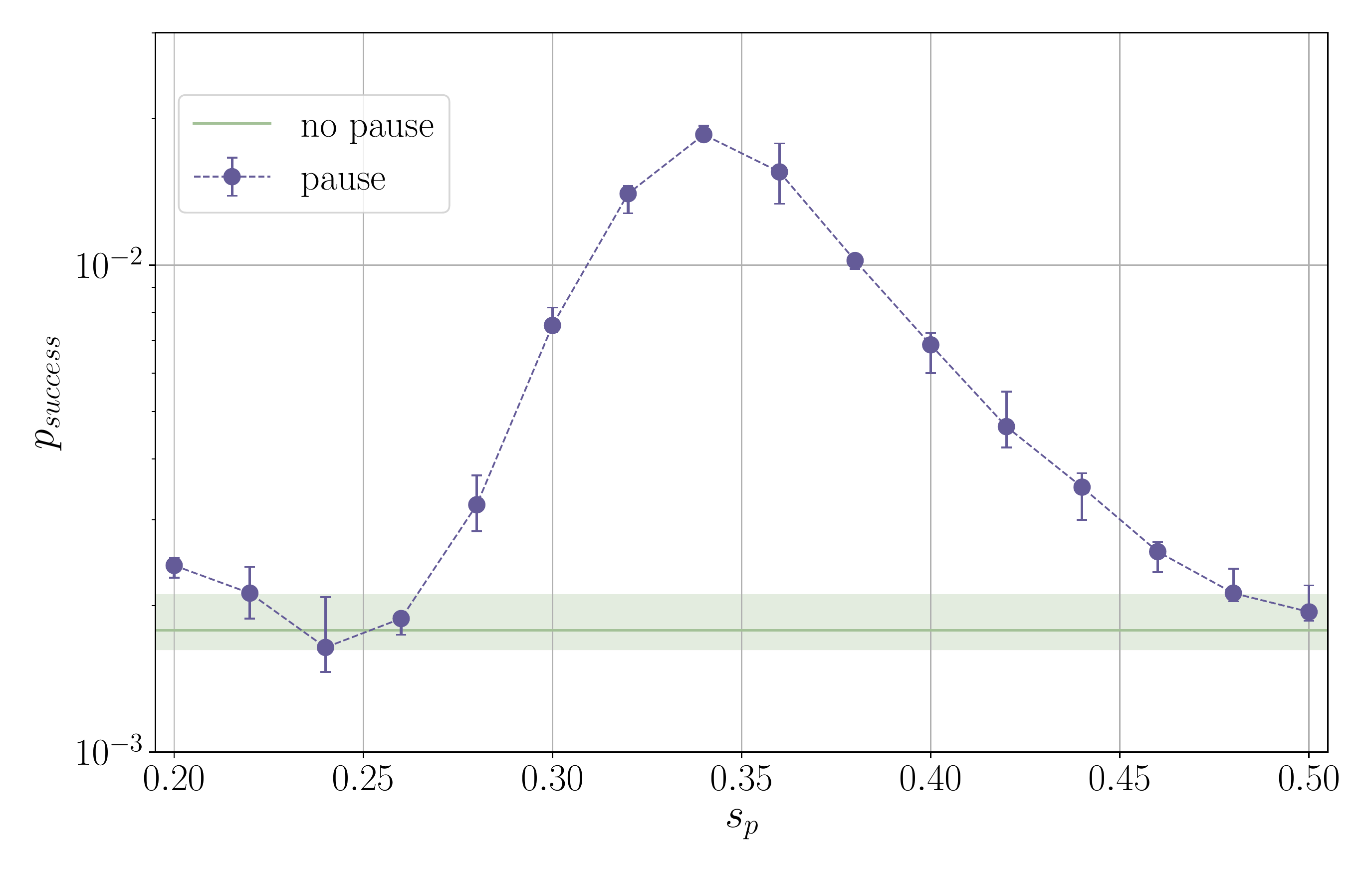}
  \caption{{\bf Improvement of $p_{success}$ with a pause.} Success probability for an ensemble of 45 instances.
  $|J_F|=1.6$ and 1 $\mu$s anneal are used. Pause duration is $100~\mu$s.
  The horizontal line shows the baseline, i.e., no-pause results.
  Each data point represents the results when introducing a pause at the location $s_p$.  At the optimal pause locations, an improvement of about an order of magnitude in $p_{success}$ is obtained.
  }
  \label{fig:pause_improv}
\end{figure}

After establishing the baseline with the no pause schedule, we introduce a mid-anneal pause. A pause can be placed at any point in the anneal, i.e., $s_p \in [0, 1]$. 
Our results show that,
like for native problems, the probability of success significantly improves when 
pausing within a specific region that is consistent across problem instances. The optimal pause location is between  $0.3\sim 0.4$, in the same range as the optimal location for the native
problems studied in Ref.~\cite{Marshall19_Pausing}. Fig.~\ref{fig:pause_improv} shows this improvement for an ensemble of 45 instances, with a pause of length $t_p = 100 \mu$s and $|J_F|=1.6$ (the optimal $|J_F|$ for the no pause case).
When we examine how the optimal pause location is affected by $|J_F|$, we will see that the peak in $p_{success}$ 
moves earlier with increasing $|J_F|$, but remains in this range.
We also find that for instances that were unsolved in the baseline (no pause) runs, a solution is 
often found after introducing an appropriate pause.  See Appendix~\ref{appendix:unsolved_cases} for  statistics on such cases.
These findings will be given theoretical and numerical support in Sec.~\ref{sec:pictures}.

\begin{figure*}[ht]
  \includegraphics[width=\linewidth]{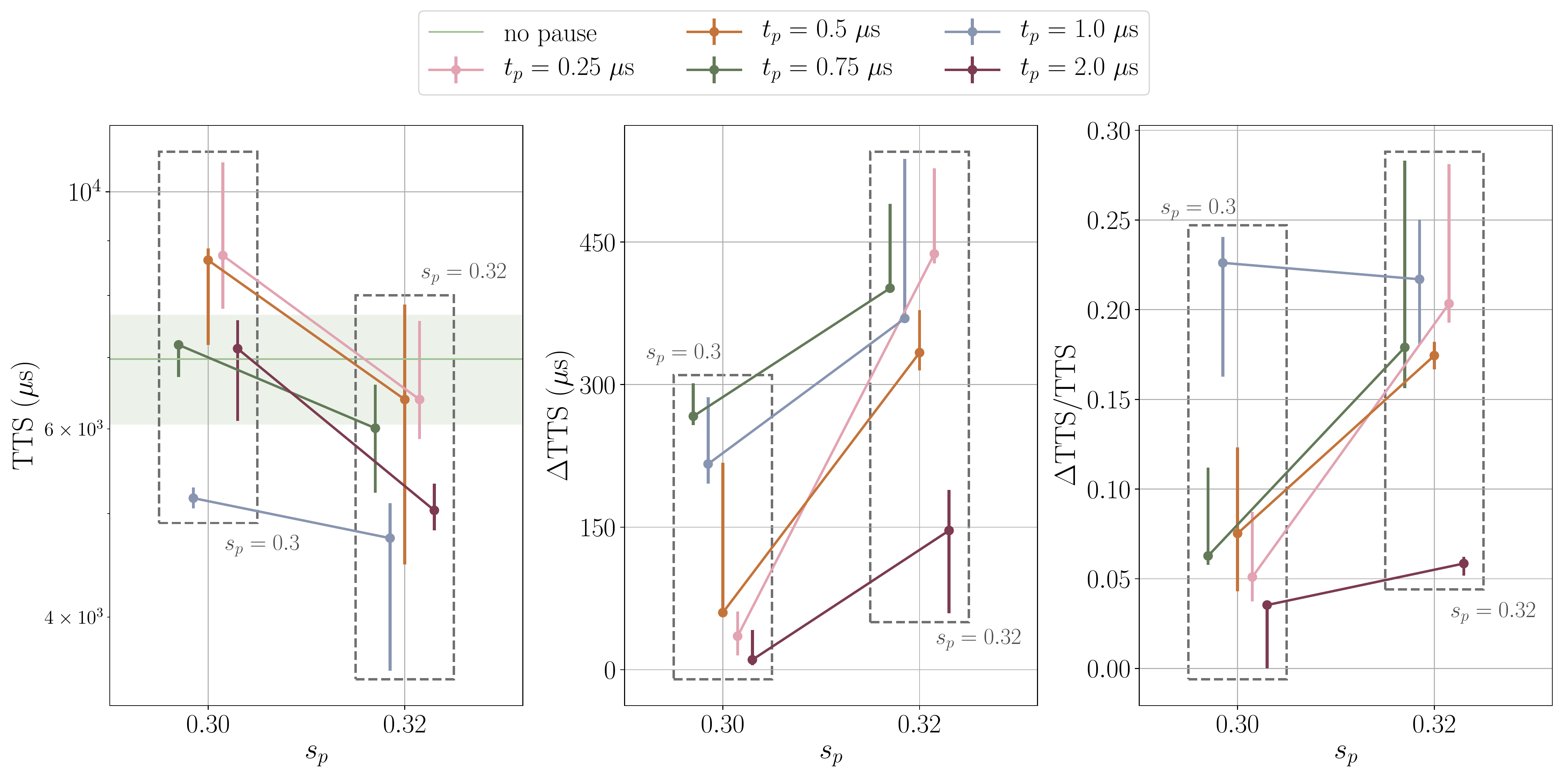}
  \caption{{\bf Effect of pause duration on TTS.}
  {\bf Left}: With pause duration of $\{0.25, 0.5, 0.75, 1, 2\}~\mu$s, and $|J_F|$=1.8, the median TTS for an ensemble of 45 instances is shown for pause locations $s_p=0.3$ and $s_p=0.32$. 
  The reference (horizontal line and band for median and 35 and 65 percentiles, respectively) is the no-pausing case with parameters optimal for TTS: $t_a=1\mu$s, and $|J_F|$=1.6. Data points show the median, with error bars at the $35^{th}$ and $65^{th}$ percentiles, after performing $10^5$ bootstraps over the set of instances.
  {\bf Center}: Instance-wise absolute improvement in TTS (in $\mu$s). $\Delta$TTS represents the reduction in TTS accomplished by introducing a short pause at an optimal location $s_p$. A positive $\Delta$TTS indicates that the TTS is reduced (improved) by the introduction of the pause. $\Delta$TTS is calculated by subtracting the TTS for the pause case with with $|J_F| = 1.8$ from that of the no pause case with $|J_F|=1.6$ (the optimal $|J_F|$ for each case). Data points are the median, error bars are $35^{th}$ and $65^{th}$ percentile obtained from $10^5$ bootstraps over 45 instances. 
  {\bf Right}: Instance-wise improvement ratio $\Delta$TTS$/$TTS. Data points are staggered along the $s_p$ axis for readability.
 Note that the errorbars in all panels are 35 and 65 percentiles of the bootstrap samples, thus indicates the uncertainty in the median values reported, instead of the median value of the 35 and 65 percentiles in the instance ensemble.
 }
  \label{fig:tts_improv}
\end{figure*}

As in Ref.~\cite{Marshall19_Pausing},
the probability of success
grows monotonically as the pause duration increases
in the range $t_p \in [0.25, 100] \mu$s (not shown). With respect to expected time-to-solution (TTS), longer duration can cancel out improvements due to increased probability of success. We were able to locate a sweet spot in pause duration for the various TTS metrics (Sec.~\ref{sec:metrics}) with pause durations of $t_p = 0.75$ or $t_p = 1.0$ (Fig.~\ref{fig:tts_improv}) at pause locations $s_p = 0.30$ or $s_p = 0.32$. We now discuss these results in more detail. 

The results of Fig.~\ref{fig:tts_improv} demonstrate that a properly placed pause of certain duration leads to statistically significant improvement in the various TTS metrics on our ensemble of \mst instances.
After sparsely sweeping through a range of parameters (not shown) 
we found that the parameter ranges $t_p \in [0.25, 2] \mu$s, $s_p = 0.3$ and $s_p=0.32$, and $|J_F|=1.8$ deserved particularly attention.
The three panels of Fig.~\ref{fig:tts_improv} correspond to the three metrics of Sec.~\ref{sec:metrics}: 1) the median TTS across the ensemble; 2) the instance-wise difference $\Delta$TTS, taking the median of this difference across the ensemble; 3) the instance-wise \emph{relative} difference $\Delta$TTS/TTS, taking the median of this difference across the ensemble. The ``median of the difference'' of the two latter metrics can be quite different from the ``difference in median''.
Since the magnitude of the TTS across our instances ranges over a few orders of magnitude, the instance-wise \emph{relative} difference $\Delta$TTS/TTS can be quite different from the instance-wise difference $\Delta$TTS.
While several of the pause schedules are better than the baseline according to every metric we use, others only do better in some of the metrics. 
The magnitude of the improvement, as well as the optimal pause location and duration, can vary significantly depending on the metric.

The left panel of Fig.~\ref{fig:tts_improv} shows TTS for the ensemble in the above narrowed parameter range.  Plotted as a horizontal line is the baseline (no pause case) at its optimal $|J_F|=1.6$. 
At both pause locations $s_p = 0.3$ and $s_p=0.32$, a pause duration $t_p=1 \mu$s is optimal on the ensemble of 45 instances. 
While at $s_p=0.3$, only the $t_p=1~\mu$s case beats the baseline,
at $s_p=0.32$, the TTS for all values of $t_p \in [0.25, 2]~\mu$s is consistently lower (better) than that of the baseline. 
(See Fig.~\ref{fig:pause_duration} in Appendix~\ref{appendix:more} for the the corresponding $p_{success}$.)

The center panel in Fig.~\ref{fig:tts_improv} shows the median instance-wise  difference $\Delta$TTS for the ensemble of 45 instances. All the data points and their respective error bars are above 0, indicating that
pausing provides a statistically significant improvement when the pause parameters are in the studied range with $s_p = \{0.3, 0.32\}$ and $t_p \in [0.25, 2]~\mu$s. 

For example, while the median TTS of the ensemble is better for the baseline case than for the pause schedule with $s_p=0.3$, $t_p=0.25$, this pause schedule did better than the baseline on more than half of the 45 instances, leading to a positive $\Delta$TTS. These two metrics provide different information about the strengths of each method.

The right panel of Fig.~\ref{fig:tts_improv} represents the instance-wise relative improvement in TTS, 
that is, each instance-wise improvement is divided by the corresponding baseline no pause TTS for that particular instance, and then the median {over the ensemble of instances} of this set of values is calculated.
We find that for a pause duration of $t_p=1\mu$s the median relative improvement holds an optimal value $\sim 0.22$. 
This pause duration was not the optimal for the absolute improvement shown in the middle panel, giving somewhat lower values of $\Delta$TTS than a pause duration of $0.75\mu$s. This `change of order' occurs whenever the following condition is met:
\begin{equation}
    \frac{\text{TTS}_j\text{(base)}}{\text{TTS}_k\text{(base)}} > \frac{\Delta\text{TTS}_j(t_p, s_p)}{\Delta\text{TTS}_k(t'_p, s'_p)},
\end{equation}
where $j$ is the instance where the median of $\Delta$TTS/TTS($t_p$, $s_p$) occurs, and $k$ the instance where the median of $\Delta$TTS/TTS($t'_p$, $s'_p$) does.
We examine in more detail the four best data points with respect to the $\Delta$TTS/TTS metric, those with $s_p \in \{0.30, 0.32\}$ and $t_p \in \{0.75, 1.0 \}\mu$s.

We first look at the absolute improvement.
At pause location $s_p = 0.3$, the median improvement for pause durations $t_p=1$ and $0.75~\mu$s were $216$ and $266$ $\mu$s, respectively.
At $s_p =0.32$, it is $369$ and $401$ $\mu$s, respectively, all the same order of magnitude.
Consider the four instances that yield these four values. Their baseline no pause TTS values vary considerably, being $897$, 5003, $2738$, and $25581$ respectively.
The substantially longer baseline TTS for the instances that are the median in each of the $1\mu$s cases than those in the $0.75\mu$s cases (5x at 0.3 and 10x at 0.32) suggests
that the $1~\mu$s pause will perform better than the $0.75~\mu$s pause under the relative difference metric. 
(This is not certain because the median in the two metrics may correspond to different instances.)

We now look at the relative improvement.
Compared to how it did with respect to the $\Delta$TTS metric, the $0.75\mu$s pause did much worse than expected relative to the other pause durations. For all four cases with parameters $s_p \in \{0.30, 0.32\}$ and $t_p \in \{0.75, 1.0 \}\mu$s, many more instances' relative performance improved with a pause than were hurt by a pause (See Tables \ref{tab:sp3} and \ref{tab:sp32}). On the other hand, for pauses at $s_p=0.3$
the median benefit over the instances for which a pause helped was less than the median amount of harm caused by a pause over the instances in which a pause hurt. This difference was much more pronounced for the $s_p = 0.30$ $t_p = 0.75$ case, with the median harm over 5 times that of the median benefit, compared to the other case where the ratio was less than $2$.
At $s_p=0.32$, the median benefit is larger than or the same as the median harm.

(In all cases, there are 3 instances that were not solved with or without a pause, hence are not included here.)

\begin{table}[ht]
\begin{center}
\begin{tabular}{| c | c | c |}
\hline
 $s_p=0.3$ & No. instances & median $\Delta$TTS$/$TTS \\
 \hline
$t_p=1$ & &\\
 \hline
  pause hurts & 15 &  -0.6666\\  
\hline
  pause helps & 27 &  0.3960 \\
\hline
$t_p=0.75$ & &\\
 \hline
  pause hurts & 14 &	{\textbf{-1.1875}} \\  
\hline
  pause helps & 28  &  0.2522 \\
\hline
\end{tabular}
\caption{$s_p=0.3$}
\label{tab:sp3}
\vskip .75pc
\begin{tabular}{| c | c | c |}
\hline
$s_p=0.32$ & No. instances & median $\Delta$TTS$/$TTS \\ 
 \hline
$t_p=1$ & &\\
 \hline
  pause hurts & 12 & -0.1993 \\  
\hline
  pause helps & 30 & 0.3467 \\
\hline
$t_p=0.75$ & &\\
 \hline
  pause hurts & 13 & -0.3602 \\  
\hline
  pause helps & 29 & 0.3879 \\
\hline
\end{tabular}
\caption{$s_p=0.32$}
\label{tab:sp32}
\end{center}
\end{table}

When interpreting these results, it is worth keeping in mind that with the exponential dependence of the TTS on the probability of solution, long TTS values are subject to much greater statistical fluctuations than shorter TTS values.

\subsection{Shift in optimal pause location with $|J_F|$}
\label{sec:shift}

\begin{figure*}[ht]
  \includegraphics[width=0.7\linewidth]{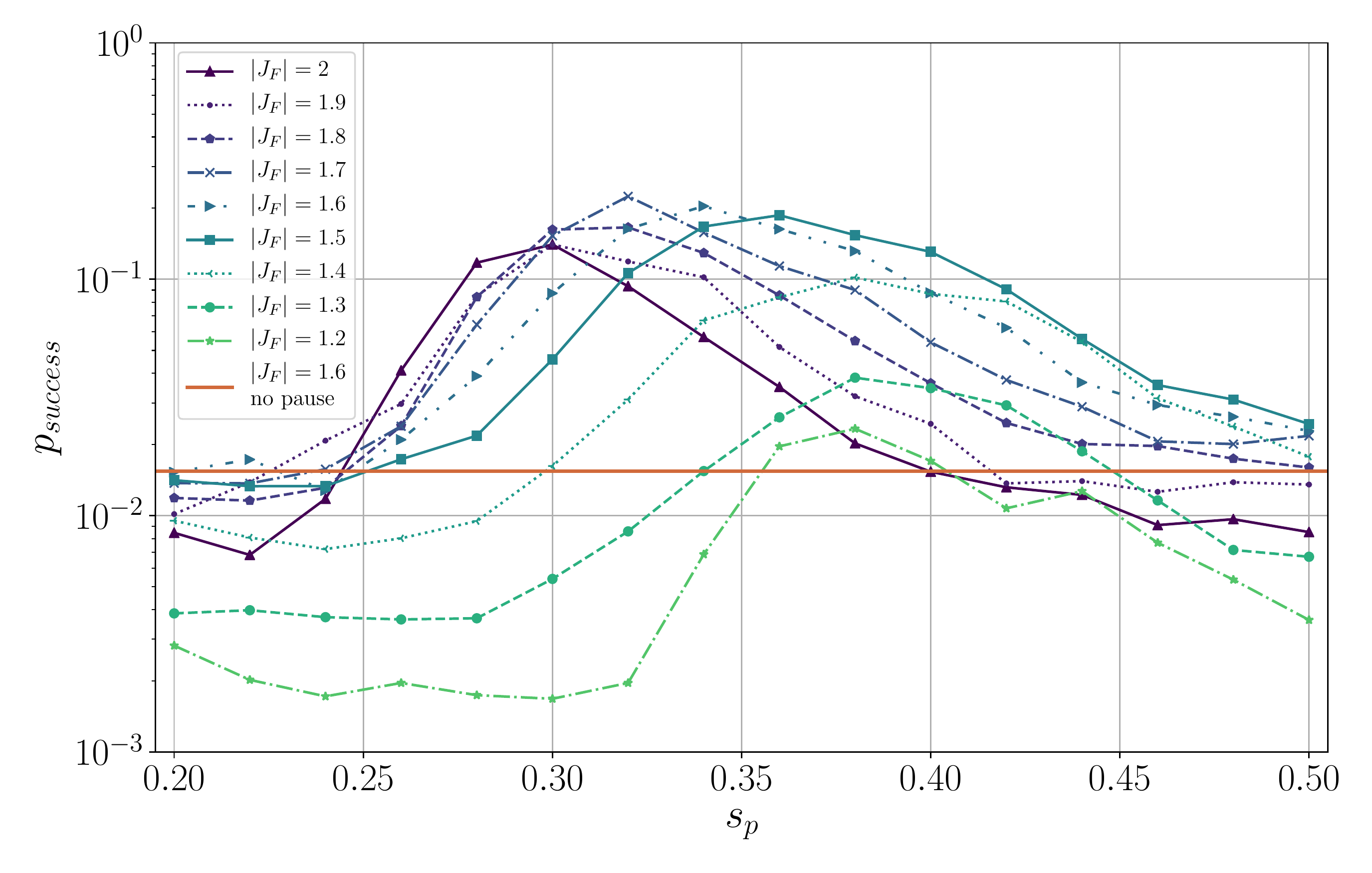}
  \includegraphics[width=.2\linewidth]{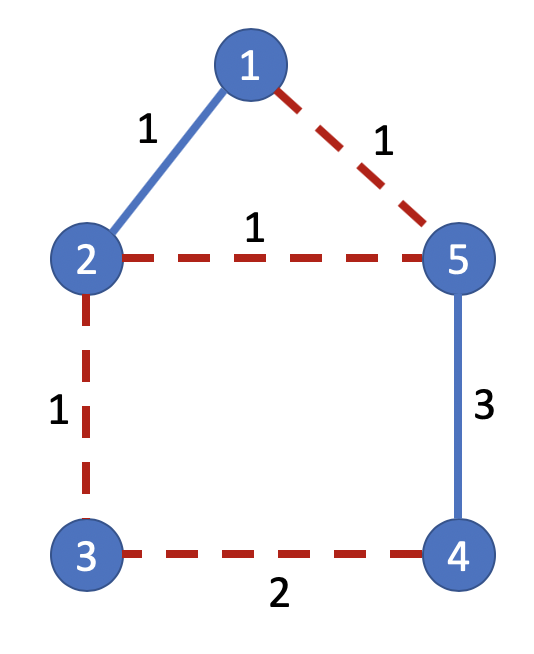}
  \includegraphics[width=0.8\linewidth]{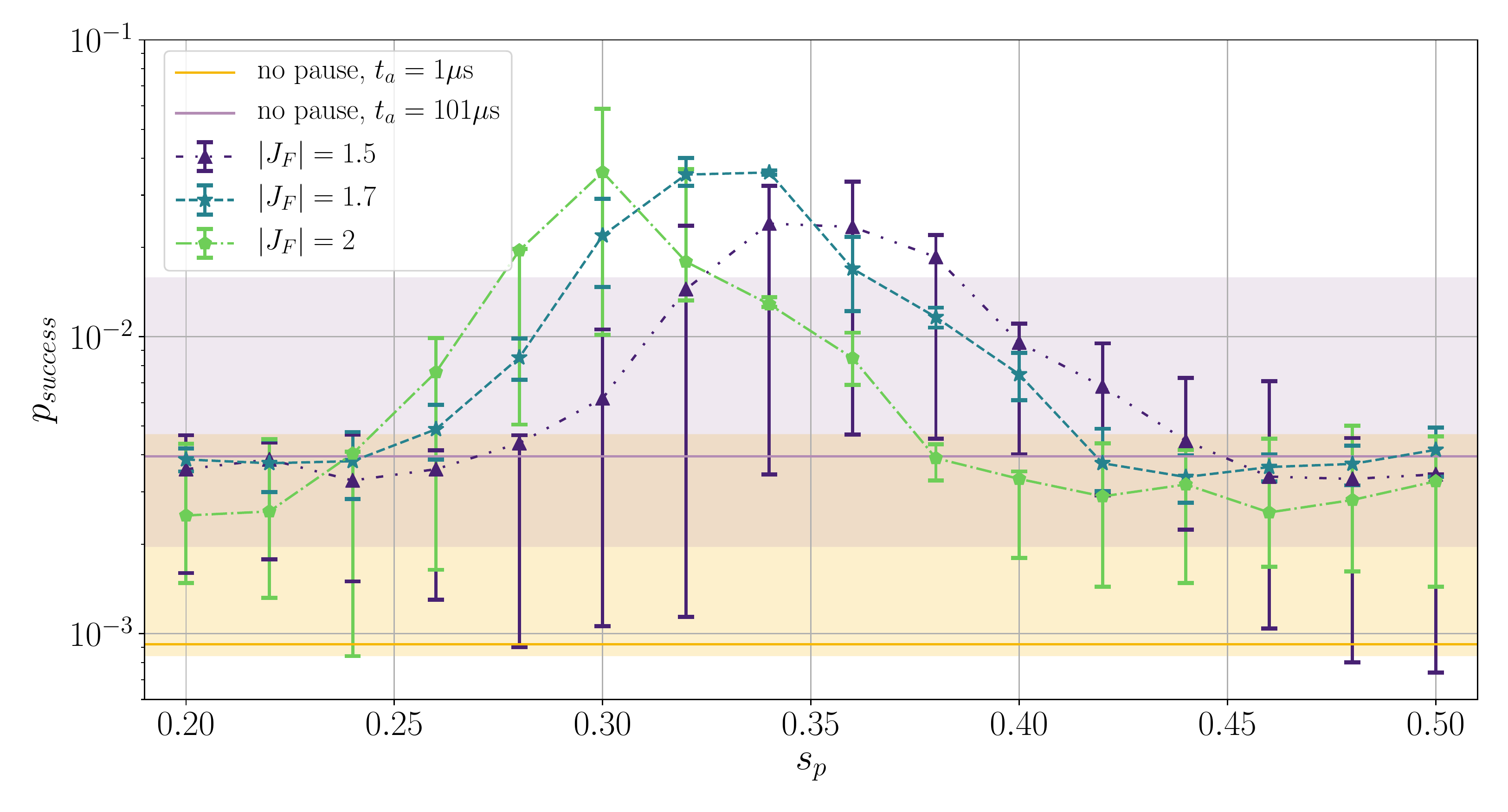}
\caption{{\bf Shift of optimal pause location with $|J_F|$.}
{\bf Top-Left}: Probability of success versus the annealing pause location for the demo instance.
The anneal was performed with 1 $\mu$s anneal time, 100 $\mu$s pause.  A monotonic shift in the peak location with $|J_F|$ is observed. The horizontal curve corresponds to the no pause, $|J_F|=1.6$, and anneal time of 1 $\mu$s.
The reason for lower success probability for $|J_F|=$1.2 is detailed in Sec.~\ref{sec:shift}.
{\bf Bottom}: Probability of success for an ensemble of $9$ instances of $n=5$ with a pause duration $t_p=100~\mu$s, and $t_a=1\mu$s. 
The horizontal lines (for median) and bands (for 35 to 65 percentiles) are baseline results with no pause, $|J_F|=1.6$:  Blue (lower) line/band: $t_a = 1~\mu$s; orange (lower) line/band: $t_a = 101\mu$s.
  }
  \label{fig:demo_shift}
\end{figure*}

One interesting new avenue that opens up with the study of embedded problems is how the value of $|J_F|$ affects the benefits and effects of pausing. As previously discussed, the $p_{success}$ vs $s_p$ curve typically shows a peak around an optimal pause location and is mostly flat far away from it (like in Fig.~\ref{fig:pause_improv}). 
We have also seen in Fig.~\ref{fig:vary_Jf} that without a pause, the value of $|J_F|$ affects $p_{success}$. 
For the pausing case, when $|J_F|$ increases,
not only does the height of the peak change with~$|J_F|$, but its position shifts as well, moving earlier in the anneal.
The top panel of Fig.~\ref{fig:demo_shift} shows this shift for a demo instance and a wide range of $|J_F|$, with the horizontal axis spanning the range of pause locations where the peak in $p_{success}$ is found.

Such clear shifting is found in many instances, and results in a shift in the behavior of the whole instance ensemble, as shown in the bottom panel of Fig.~\ref{fig:demo_shift}.
For figure clarity, pausing results for just three values of $|J_F|$ are shown.  
The shift is consistent over all $|J_F|$ values we examined (in $[1.2, 2]$); see Fig.~\ref{fig:ensemble_shift_new} in Appendix~\ref{appendix:more} for additional results.

The success probability for smaller $|J_F|$ values, like $1.2,~1.3$, even away from the peak, is clearly lower than for larger $|J_F|$ values  (This holds true for the ensemble of instances, but some individual outliers have been found, with a high $p_{success}$ for smaller values of $|J_F|$. Fig.~\ref{fig:more_instances} in the Appendix shows some examples). The reason is that when the ferromagnetic coupling is not very strong compared to the problem couplings, it is more likely that the low-lying energy states are densely populated by states with an inconsistent vertex model (i.e. when not all the qubits are aligned and hence are no longer acting as a single variable).  
Accordingly, even when the annealer is doing well at finding the ground state or a low-lying state, such outcome do not correspond to a valid solution of the original problem.  Indeed, by applying simulated annealing to solve the embedded problem (which is too large to diagonalize exactly), we verified that in the range of $|J_F|$ we used,  the ratio of inconsistent-vertex -model-state in the ground/low-lying states is significantly higher for $|J_F|=1.2,~1.3$ than that for $1.4$ and above.

The mechanism for why the optimal pause location typically shifts toward earlier in the anneal with $|J_F|$ fits our theoretical understanding, which is laid out in section~\ref{subsubsec:jf_effect}.

\subsection{Help of Partial gauges}
\label{sec:partial_gauges}
\begin{figure}[ht!]
  \includegraphics[width=\linewidth]{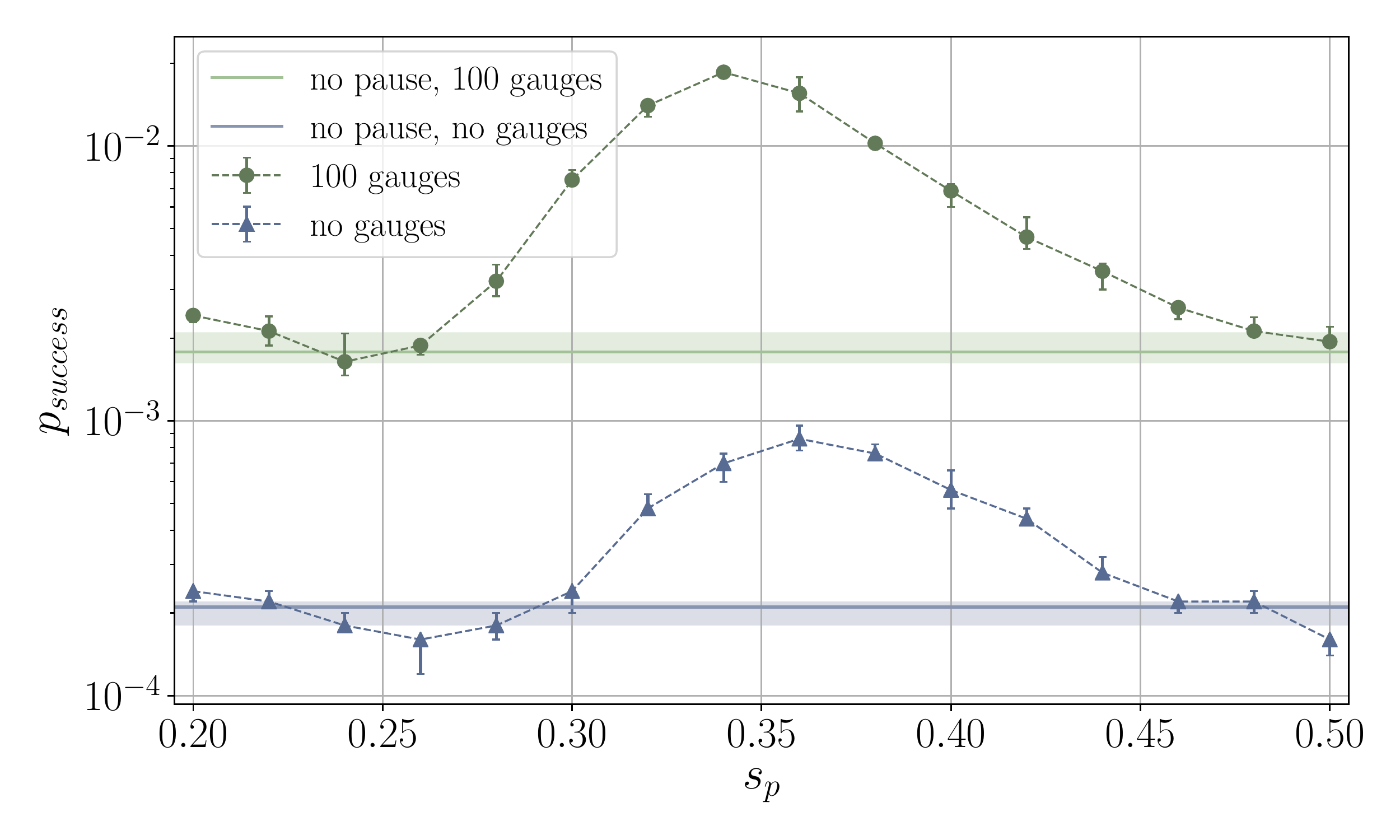}
  \includegraphics[width=\linewidth]{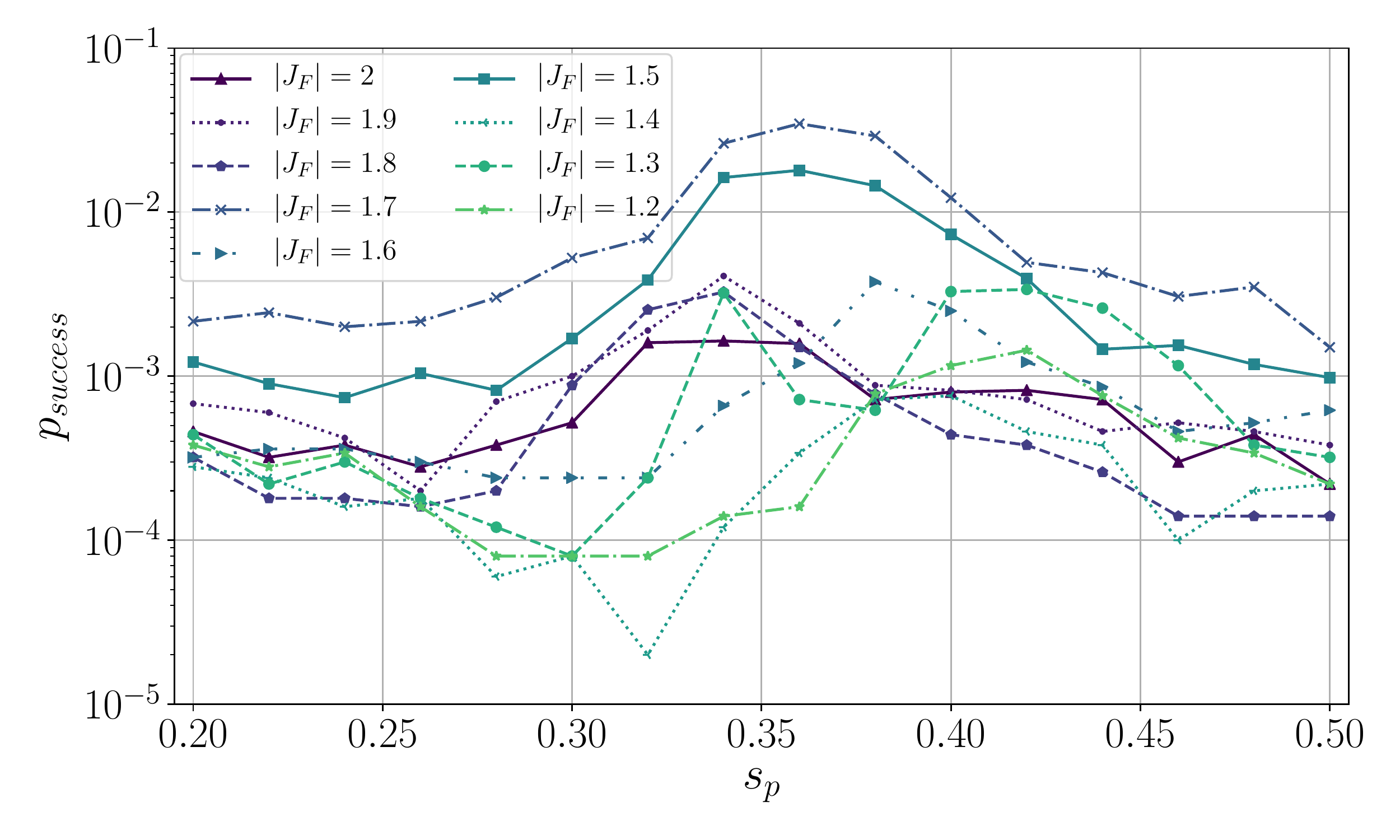}
  \includegraphics[width=\linewidth]{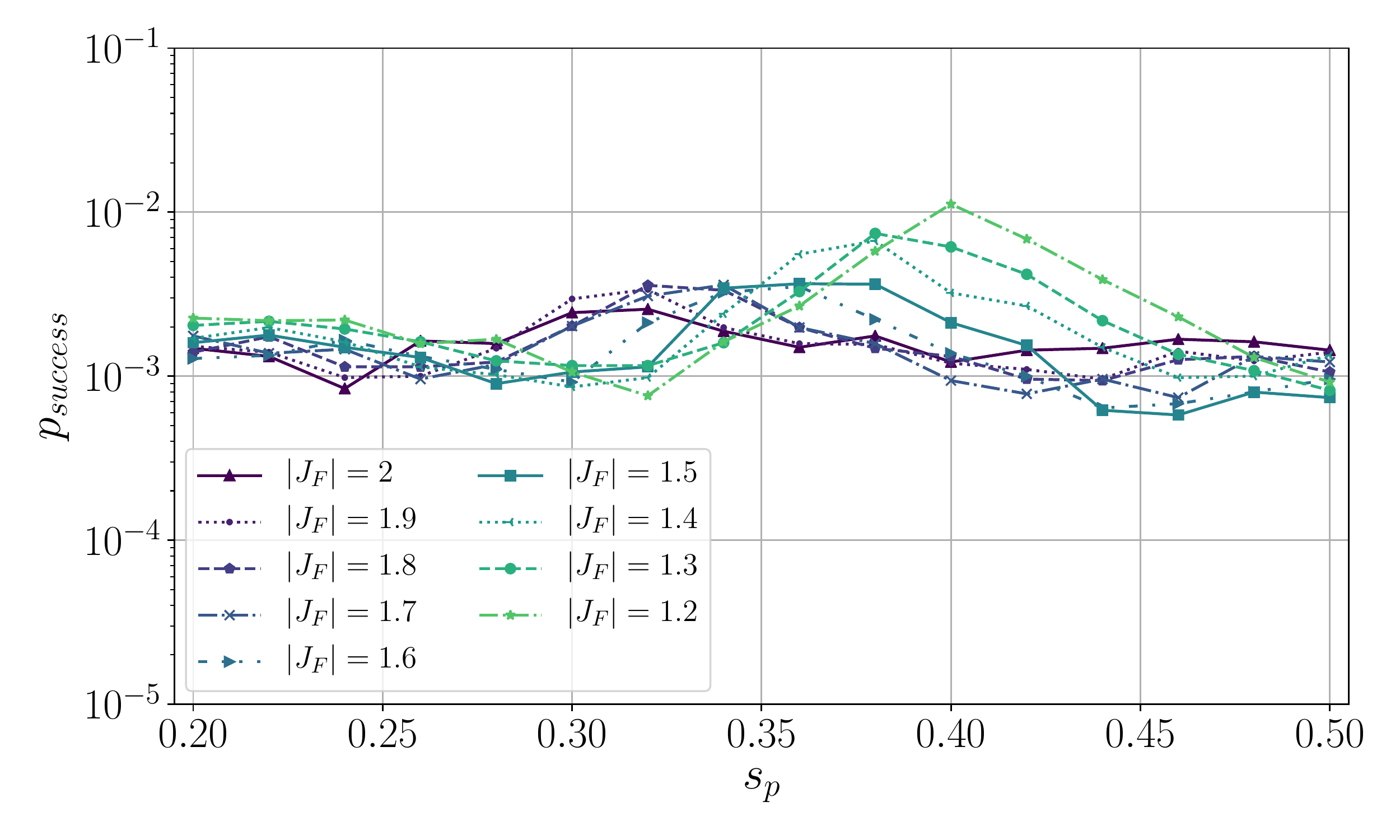}
  \caption{{\bf Effect of partial gauges}.
  {\bf Top:} partial gauges help boost average success probability for a set of 117 instances.  $|J_F|=1.6$ is used for all data shown.
  The baseline case without pausing is shown for reference: the median is shown as the horizontal lines and the 35 and 65 percentiles are shown as half-transparent bands.  Blue (lower) band is for no gauges and orange (upper) band is for 100 gauges applied.
  {\bf Middle:} Effect of $|J_F|$ on $P_{succ}$ for a single instance. $t_a = 1 \mu$s, a pause of duration 100 $\mu$s is applied.
  No  gauge transformations is performed. 
  {\bf Bottom:} The same instance and parameters as in the Middle, but with 100 partial gauges applied.  Partial gauges helped suppress the variance, and revealed the peak shift with $|J_F|$.
  }
  \label{fig:gauge_cmp}
\end{figure}

We developed a \emph{partial gauge transformation} technique that significantly improved the success probability, and enabled the confirmation of the peak shift.  

Gauge averaging is a technique commonly used to alleviate the effect that intrinsic biases on the local fields and couplers can have on the data obtained from a quantum annealer~\cite{Boixo2013experimental}. It can help improve statistics and lead to less noisy results and improved $p_{success}$ and TTS.
A gauge transformation starts with assigning a random sequence $a_j \in \{\pm 1\}$ to re-define the basis for each qubit, $\tilde Z_j = a_j Z_j$.  If we accordingly adjust the local field and couplers such that
\begin{align*}
\tilde J_{ij} &= a_i a_j J_{ij} \nonumber \\
\tilde h_i &= a_i h_i\;,
\end{align*}
then the resulting Hamiltonian has the same energy spectrum as the original one.  This Hamiltonian is run on the annealer and the output bit string is transformed back using the same $a_j$'s.  By performing multiple gauge transformations and averaging results over them, biases that stem from, for example, a qubit having a slight preference to aligning in one direction over the opposite, can be suppressed.  

When $J_{ij} \in [-1, 1]$, it is straightforward to apply gauges. For our embedded problems, however, we are making use of D-Wave’s extended $J$ range, allowing $J_{ij} \in [-2, 1]$. The extended range discourages the breaking of vertex models during annealing thanks to the stronger ferromagnetic couplings between physical qubits representing the same logical variable, but it also impedes the use of standard gauges, since any couplings in the range $[-2, -1)$ cannot change sign.

Our partial gauge method circumvents this issue by only applying the gauge transformation on the couplings within the interval $[-1, 1]$. Because the extended range is exclusively used on the vertex models in our problems, the partial gauge on the embedded problem is equivalent to applying a general gauge to the logical problem before embedding.

In a previous study~\cite{kim2019}, the boost in $p_{success}$ by pausing is observed for a family of embedded problems, but no relation between the optimal pausing location and $J_F$ was observed. 
In our study, with the help of partial gauge transformation, the variance in the annealing output is significantly suppressed, resulted in the revelation of the shift
of the peak in Sec.~\ref{sec:shift}. This improvement of the variance is seen in the top panel (note log scale) and by comparing
the middle and bottom panels of Fig.~\ref{fig:gauge_cmp}.

Another benefit was a remarkable increase in $p_{success}$ for hard problems.
Usually, we don’t expect $p_{success}$ to change significantly from gauge averaging, because solving the problem without gauge transformations is just applying one gauge, which typically will be near average instead of an outlier. 
But the success probability is lower bounded by zero, and when
problems like the ones we are solving here are difficult for the solver the typical empirical $p_{success}$ is zero or very close to zero.
The existence of such a lower bound explains the significant 
benefit in applying gauges: even if we get a bad gauge, $p_{success}$ cannot go below 0,
while a good gauge can yield a much higher $p_{success}$.  In a number of gauges it is likely to encounter a few good gauges, bringing the average $p_{success}$ up.
The top panel of Fig.~\ref{fig:gauge_cmp} shows, for a large ensemble of instances, the improvement in success probability with 100 partial gauges is significant: about an order or magnitude higher than the results ran without gauge transformation.

The improvement in $p_{success}$ will saturate as one increases the number of gauges applied.  Fig.~\ref{fig:n4_gauges} in the Appendix shows for an $n=4$ ensemble, applying as few as 10 gauges yields similar $p_{success}$ to 100 gauges.  These results indicate that with 10 gauges we are already likely to encounter one or more positive outliers, leading to the large improvement in $p_{success}$.  As the number of gauges further increases, the effect is not as dramatic, indicating the spread in gauge quality approaches the intrinsic distribution.

The partial gauge transformation therefore enables us to extend the benefits of general gauge averaging to embedded problems. 

\section{Physical picture}
\label{sec:pictures}
In this section, we 
expand the physical picture of Ref.~\cite{Marshall19_Pausing} to embedded problems, explaining both the shift of optimal pausing location with increasing $|J_F|$ and why embedded problems, while benefiting significantly, benefit less from a pause than native problems. We provide a perturbation analysis supporting the picture, and numerical evidence on the change in minimal gap location. The picture is far from that of the adiabatic regime - pausing is effective after not at the minimum gap and diabatic and thermal effects play a significant role.

\subsection {How pausing helps}
We start with a recap of the physical picture of Ref.~\cite{Marshall19_Pausing} that 
explains the increase in success probability by introducing a pause in the middle of the annealing schedule, after the minimal energy gap. 
Recent work \cite{chen-pausing} verified 
this qualitative picture in numerical simulations, and also provided sufficient conditions under which pausing improves success probability. Loosely speaking, so long as shortly after the minimum gap the relaxation time-scale is small enough  (relative to the pause time), 
one can expect a pause to boost success probability. As discussed above, whether or not this improves the TTS is not as obvious.

We use GS and FES to refer to the ground and subspace of first excited states of the instantaneous quantum Hamiltonian.  In the rest of the section we refer to the {\it gap} as the energy gap between the GS and the FES.

At very early or late stages in the annealing, only one Hamiltonian---either the driver $H_X$ or the problem $H_C$---dominates. Since both the problem Hamiltonian and the driving Hamiltonian are classical when acting alone, dynamics in these regions are almost classical.  Because the temperature $T$ is much lower relative to the energy scale, thermal relaxation rates remain slow.

In the middle of the anneal, when the scales of $H_X$ and $H_C$ are comparable, the system dynamics is determined by the interplay of the energy gap, non-adiabaticity (annealing speed relative to the gap), and thermalization.
In this region we expect significant population loss from the ground state to excited states.
In particular, when the gap is small enough, approximate instantaneous thermalization may occur, populating excited states. This region is also where non-adiabiatic transitions are expected to be largest.

\begin{figure}
\includegraphics[width=\linewidth]{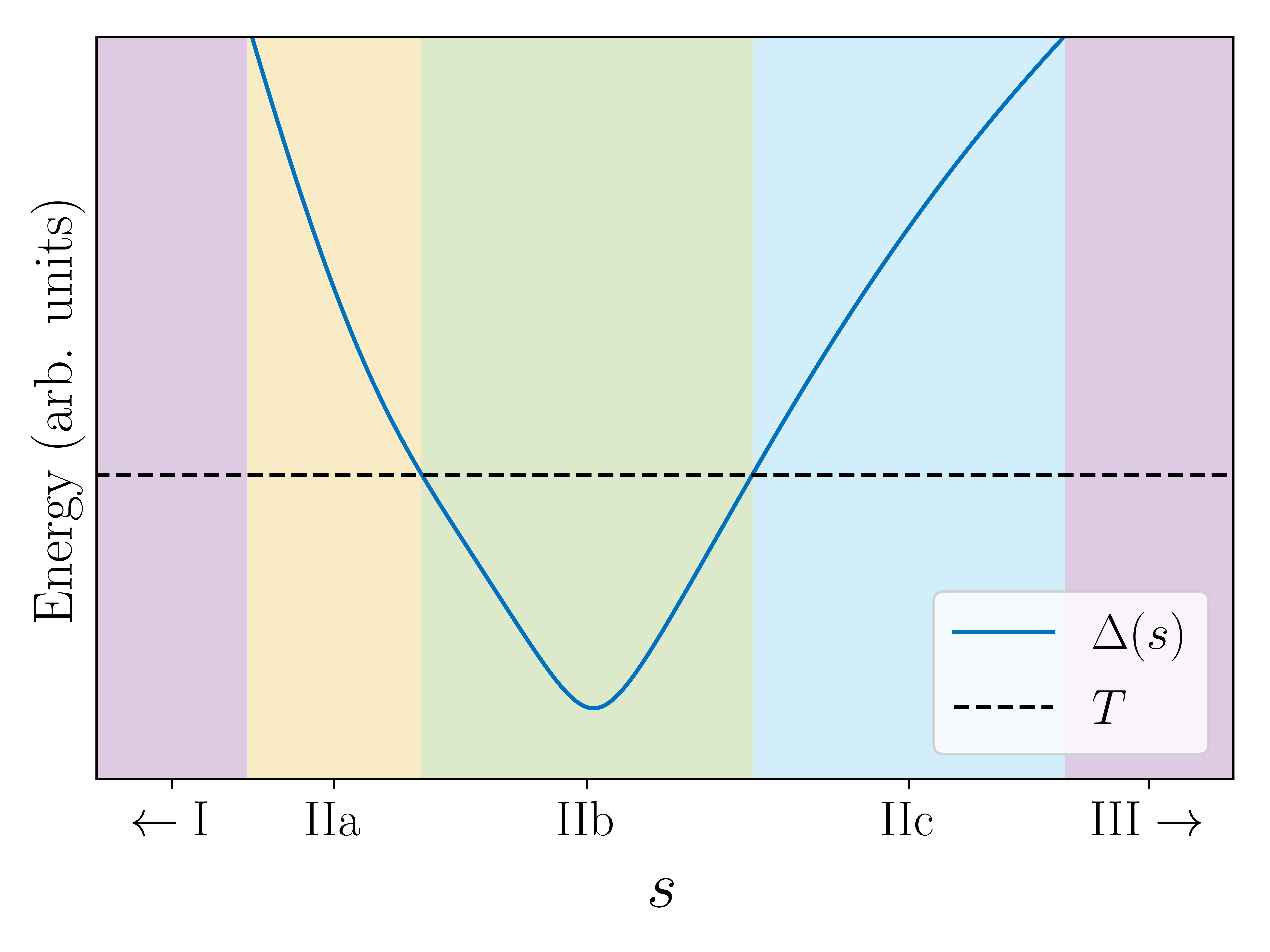}
  \caption{{\bf Cartoon diagram for the three Regimes.} Colored in purple the left and rightmost regions are the adiabatic Regimes I and III (which further extend to the left and right as indicated by the arrows). Regime 2 is further subdivided into three regions, a,b,c, as in the main text, and are determined by the instantaneous gap $\Delta$ and the temperature $T$.  In IIa we expect the system to stop behaving strictly adiabatically, and region IIb instantaneous thermalization may occur if the relaxation time scale is small enough, as well as non-adiabatic transitions. In IIc, a pause may help to repopulate the GS. This should be thought of as an approximate picture of what occurs, to aid the reader. In reality the transitions between these regions will of course not be sharp and defined by a single point during the evolution.}
  \label{fig:cartoon}
\end{figure}

We thus distinguish three different regimes in the anneal, as described below and illustrated by the cartoon schematic 
in Fig.~\ref{fig:cartoon}.

{\bf Regime I}: $||A(s)H_X||\gg ||B(s)H_C||$. The instantaneous Hamiltonian is mainly $H_X$, and its energy scale is much larger than the temperature, $T$. The system stays in the ground state of $H_X$.  
{\bf Regime II}: $||A(s)H_X|| \sim ||B(s)H_C||$, and their energy scale is comparable to the temperature.  Both thermal and quantum dynamics happen, and the minimal gap occurs in this region.
Thermalization and non-adiabaticity are both contributing to populating FES (compared to the case of zero-temperature adiabatic evolution in which all population is in the GS).

As the anneal goes on in this regime, it sequentially goes through the following stages:
\begin{enumerate}[label=\alph*]
\item Gap approaching temperature, system leaving adiabatic regime, but transitions (non-adiabatic and thermal) may still be relatively slow compared to the system evolution.

\item 
\label{stage:critical}
Gap is near its minimum and is much smaller than the temperature --- thermalization happens almost instantaneously and the system is near the thermal equilibrium state. 
Quantum non-adiabatic effects could be strong enough to increase the population of the FES beyond its magnitude at thermal equilibrium. 

\item
\label{stage:fixed}
Gap is larger than temperature, non-adiabaticity is weak.
System may still approximately equilibriate if given enough time (e.g. a pause), but will not fully thermalize during the standard evolution.
\end{enumerate}

At stage II~\ref{stage:critical}, the instantaneous equilibration removes the state memory from the history. The system is simply in thermal equilibrium. Due to the closeness between the GS and FES, the FES is significantly populated.

As the system enters stage II~\ref{stage:fixed} from II~\ref{stage:critical}, pausing promote betters thermalization, which could bring significant FES population in stage II~\ref{stage:fixed} down to the GS, since relative to the gap the temperature $T$ is now lower, hence boosting the success probability.

{\bf Regime III}: $||B(s)H_C|| \gg ||A(s)H_X||, ||B(s)H_C|| \gg T$, dynamics are slow,
the system simply picks up phases under $H_C$, and the population distribution is final. This is also known as the frozen region in the literature.

\subsection {How $|J_F|$ shifts the optimal pausing location earlier}
\label{subsubsec:jf_effect}

An increase in $|J_F|$ is expected to shift the minimal gap to earlier in the anneal, meaning that stage II~\ref{stage:critical} occurs earlier in the anneal, and therefore also shifting the optimal pause region II~\ref{stage:fixed} earlier.
This shift of the minimal gap can partly result from the increase in the relative norm of $H_C$, i.e. decreasing the value of $\frac{A(s)}{B(s)}\frac{\|H_X\|}{\|H_C\|}$. Similar to Ref.~\cite{Choi19}, this is akin to shifting each point earlier in the anneal.

\begin{figure}
\includegraphics[width=\linewidth]{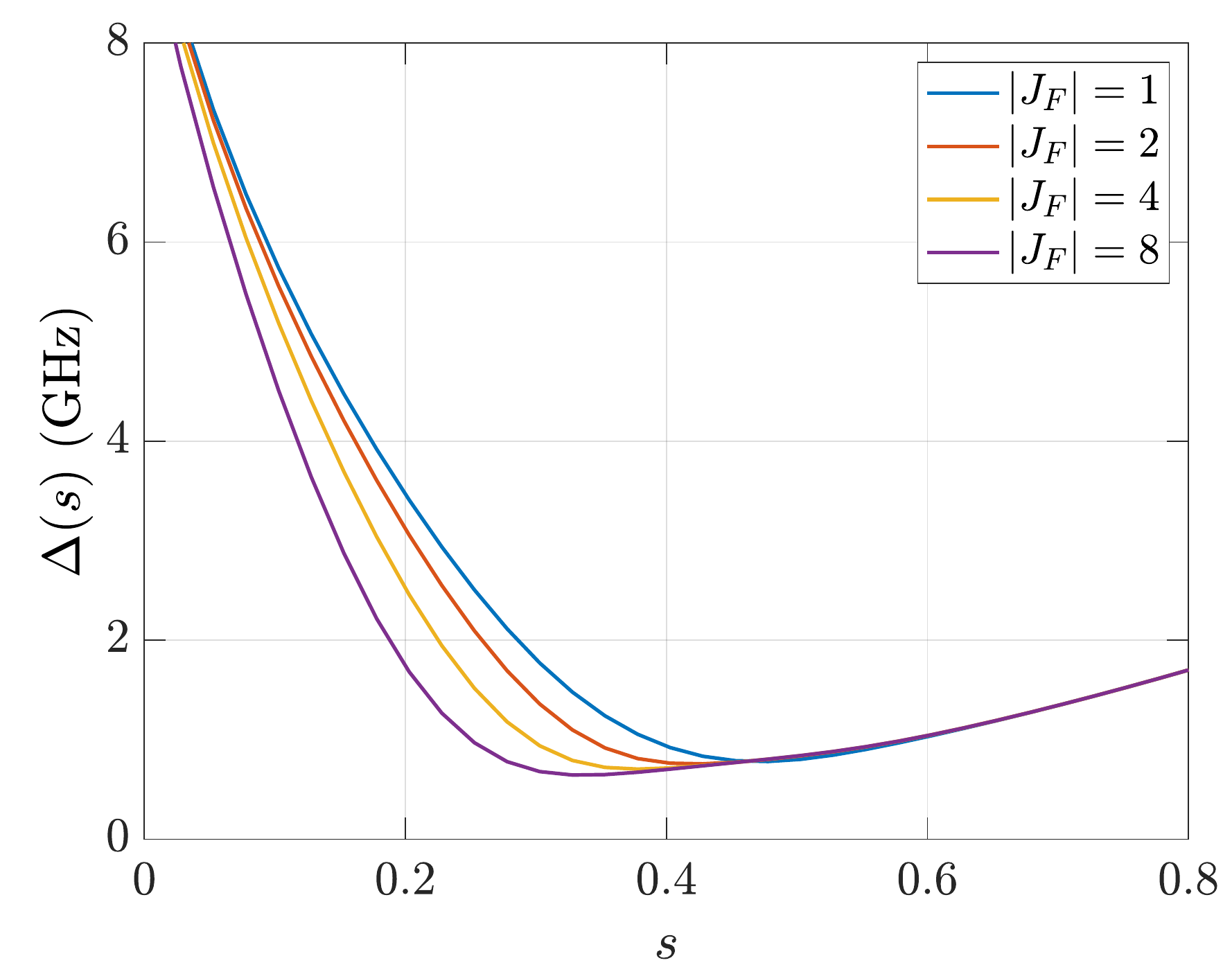}
  \caption{{\bf Shift of minimal gap with $|J_F|$}.
  Energy gap between the ground and the first excited states for the instantaneous quantum Hamiltonian during annealing for a toy problem. The logical problem is an Ising problem of a complete graph of size 3, embedded to 4 physical qubits of Chimera connectivity.  The gap is computed exactly by diagonalizing the instantaneous Hamiltonian.  As $|J_F|$ increases, the instantaneous gap closes, and the minimum gap shifts to earlier in the anneal.
  }
  \label{fig:min_gap}
\end{figure}

In Figure.~\ref{fig:min_gap}, for a small Ising problem embedded to 4 physical qubits with Chimera connectivity---which allows exact diagonalization of the instantaneous Hamiltonian---we show the change of minimal gap with $|J_F|$. Note that because the cells in Chimera are bipartite graphs, odd cycles are not native to the structure. In this small example, a triangle on 3 nodes requires minor embedding as a square on 4 qubits.
Below we provide an argument as to why the minimum gap increases in value with decreasing ferromagnetic strength.

Before providing a proof sketch for the gap increase, we mention another picture that comes into play is that an increase in $|J_F|$ can yield “clusters” (physical qubits representing the same logical qubits) with stronger internal couplings.  Changing the state in such clusters requires collective flipping of qubits, and demanding greater quantum dynamics.
Accordingly the transition from stage II~\ref{stage:critical} to II~\ref{stage:fixed} would happen earlier in the anneal.
Such a picture may also be accountable for the less dramatic increase in success rate compared to the native Ising case: the associated energy barrier may require much higher relative temperature, while pausing earlier helps, the amount it can help is limited (because it is an interplay of the three influences which are correlated in a given annealing schedule and at a given temperature).

\textbf{Proof sketch of gap scaling under $|J_F|$:}
We apply first order non-degenerate perturbation theory. Let $H(s) = H_0(s) + B(s)\lambda H_F$ with $H_0(s) = A(s)H_X + B(s)H_C +B(s)J_F H_F$ where $\lambda>0$, $J_F<0$ and $H_F$ is the ferromagnetic Hamiltonian for the vertex model.
That is, we are considering the effect of weakening the vertex model infinitesimally by decreasing $|J_F|$.
To simplify matters, assume the only vertex model is a chain of length 2. Then $H_F=\sigma_{k_1}^z \sigma_{k_2}^z$ for two qubits $k_1,k_2$.
Write $|E_i(s)\rangle$ as the instantaneous $i$th eigenstate of $H_0(s)$.
For simplicity, we drop the explicit~$s$ dependence (i.e., we will just consider~$s$ fixed at some value). Then we can always decompose our instantaneous eigenstates in the computational basis,
$|E_i\rangle = \sum_j a_j^{(i)}|z_j^L\rangle + \sum_k b_k^{(i)}|z_k^B\rangle$, where $|z_j^L\rangle$ are logical states, and $|z_k^B\rangle$ has the chain broken.
We compute the matrix elements 
\begin{equation}
    \langle E_i|H_F|E_i\rangle = \sum_j |a_j^{(i)}|^2 - \sum_k |b_k^{(i)}|^2.
\end{equation}
Note, by normalization $\sum_k |b_k^{(i)}|^2 = 1- \sum_j |a_j^{(i)}|^2$, and, denoting the logical probability $P^{(i)}_L:=\sum_j |a_j^{(i)}|^2$,
\begin{equation}
    \langle E_i|H_F|E_i\rangle = 2P_L^{(i)} - 1.
\end{equation}

\begin{figure}
\includegraphics[width=\linewidth]{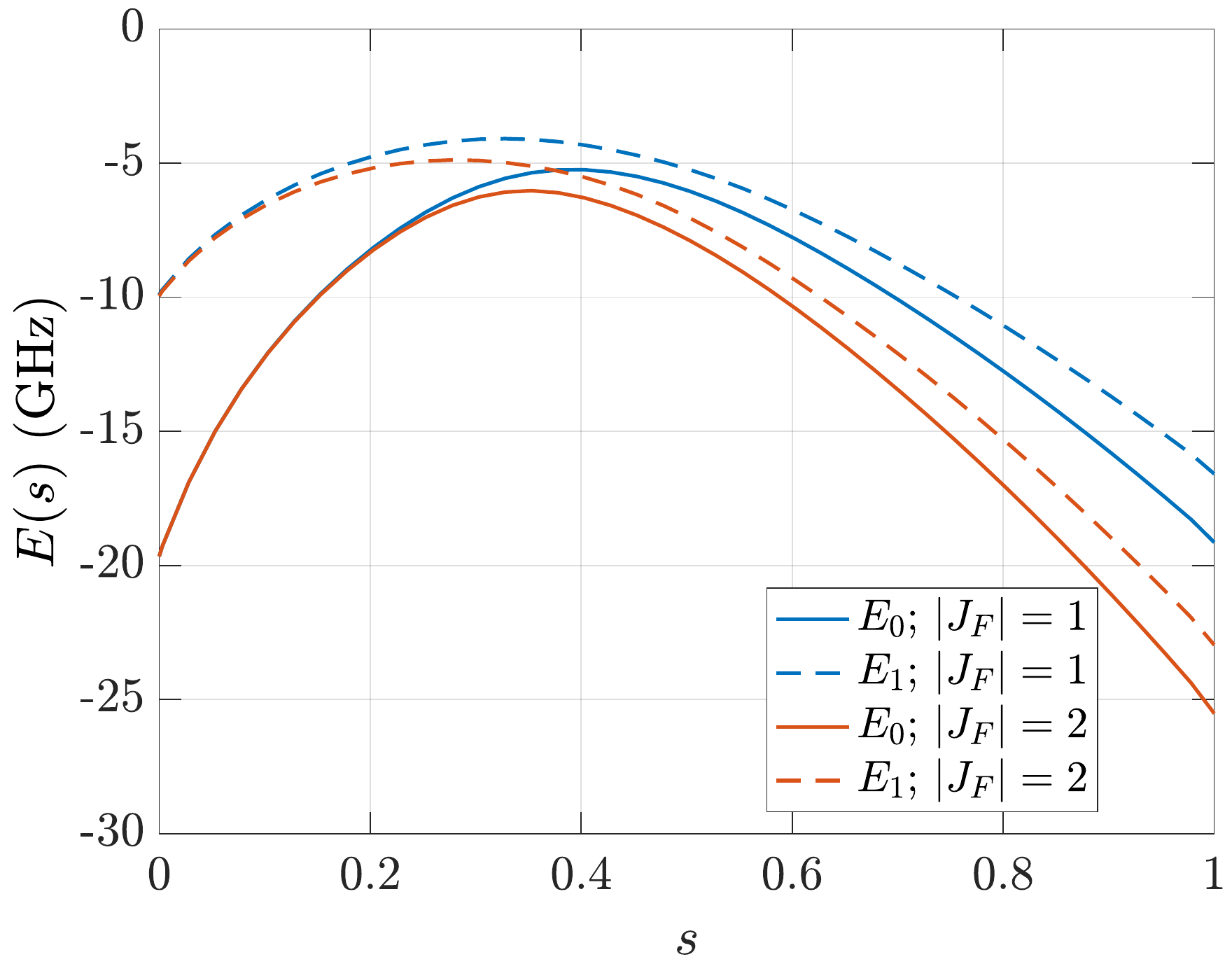}
  \caption{{\bf Energy level shift with $|J_F|$}.
  The individual energy levels show an increase in energy upon decreasing the ferromagnetic strength (i.e. the case when $\lambda>0$ in our perturbation theory). This is for a problem with a single chain of size 2 (complete graph of size 3, embedded size 4). The gap itself is shown in Fig.~\ref{fig:min_gap}.
  }
  \label{fig:gs_fes}
\end{figure}

This tells us, to first order in $\lambda>0$, that the low lying energy levels experience an increase in energy upon decreasing the ferromagnetic strength ($|J_F|\rightarrow |J_F|-\lambda$), i.e., $E_i' = E_i + B\lambda(2P_L^{(i)}-1)>E_i$, assuming that $P_L^{(i)}>1/2$. We see consistent behaviour with this picture in Fig.~\ref{fig:gs_fes} (even though this figure is not in the perturbative limit).

Now, the  gap $\Delta = E_1-E_0$ changes under $\lambda$, to first order, as
\begin{equation}
\begin{split}
    \Delta' = \Delta + B\lambda (\langle E_1|H_F|E_1\rangle - \langle E_0|H_F|E_0\rangle) \\
    = \Delta + 2B\lambda (P_L^{(1)}-P_L^{(0)}),
    \end{split}
    \label{eq:delta_pertub}
\end{equation}
which therefore increases in magnitude (at a fixed $s$) by weakening the ferromagnetic couplings, assuming $P_L^{(1)}>P_L^{(0)}$. 

Note that at the start of the anneal, $|E_0(0)\rangle = |+\rangle^{\otimes N}$, and so $P_L^{(0)}=1/2$ (in the specific case when the embedding contains just one additional qubit). At $s=0$, FESs are linear combinations containing one excitation in the $x$ eigenbasis, i.e., a single $|-\rangle$.
Consider the symmetric FES, denoted $|FE_+\rangle$, where the state of the chain is $\frac{1}{\sqrt{2}}(|-+\rangle + |+-\rangle)$ (and the other qubits are all $|+\rangle$). 
This state is entirely in the logical subspace, due to the cancelling out of the $|01\rangle$ and $|10\rangle$ terms.
When the transverse field is `strong', i.e., `near' to $s=0$ (but where the FES degeneracy is broken), by the perturbation theory we may indeed therefore expect that $\Delta' > \Delta$. We see this in Fig.~\ref{fig:min_gap}, where the strongest chain, $|J_F|=8$, has the smallest instantaneous gap. In the arbitrary chain length case, following the general expression the first line of Eq.~\eqref{eq:delta_pertub}, a similar argument applies provided $P_L^{(1)}$ is large enough relative to $P_L^{(0)}$, though the precise dependence is more complicated.

We also know that once the transverse field becomes weak relative to the problem Hamiltonian (e.g. $A/B < 1$), that $P_L^{(1)}-P_L^{(0)}\rightarrow 0$ as both the instantaneous GS and FES become close to logical states.

By interpolating between the two extremes, the above argument explains the change in gap size observed in Fig.~\ref{fig:min_gap}.

\section{CONCLUSIONS AND FUTURE WORK}
\label{sec:conclusion}

We studied how mid-anneal pauses affect performance on embedded problems
using the previously unconsidered class of degree-bounded minimum spanning tree problems. We developed a partial gauge approach that allowed us to take advantage of the extended $J$-range while also using gauges (partially), yielding significantly cleaner results and improved performance than without partial gauges, enabling us to confirm theoretical predictions. 
Our results confirm that, like 
for native problems, 
there is a region, consistent across instances, in which a pause
improves the probability of success. We further showed that the pause generally improves the time-to-solution (TTS) for these problems and evaluated the performance on three TTS-related metric.
We extended the theoretical picture of \cite{Marshall19_Pausing} to embedding problems, describing the interaction of embedding parameters with annealing parameters, thermalization, and non-adiabatic effects. 
This picture explains why the optimal pause location moves earlier in the anneal as $|J_F|$ increases and why the benefit provide by pausing, while significant, is not as great as for native problems. It generally provides
both deeper insights into the physics of these devices and pragmatic recommendations to improve performance on optimization and sampling problems.

This study suggests a number of avenues for future research. As the connectivity of quantum annealing hardware increases, as is anticipated in D-Wave's upcoming Pegasus architecture, lower embedding overhead should translate to greater benefit from pausing. Larger and more connective devices will allow larger problem sizes to be run, enabling scaling analyses. As annealing hardware becomes more flexible, a wider variety of advanced schedules become possible such as a smooth slowing down rather than a pause or annealing at different rates in different parts of the system depending on the local embedding characteristics or local problem instance structure. All of these possibilities should be explored on a variety of optimization problems as well as on the BD-MST problem class investigated here. Embedding affects sampling problems even more than optimization problems \cite{marshall2020perils}, so a study of the interplay between embedding parameters and annealing parameters should be done in that context as well. Experiments at other temperatures and with the ability to do quick quenches at arbitrary points in the anneal would give further insight in the the underlying physics. Further, given that diabatic behavior is expected to be useful even for devices that could remain adiabatic through out a run, an intriguing area for both theoretical research and hardware development is the use of engineered dissipation to support cooling in conjunction with diabatic evolution, enabling much more controlled utilization of thermalization in quantum annealers of the future.

\section{Acknowledgments}
We are grateful for support from NASA Ames Research Center, particularly the NASA Transformative Aeronautic Concepts Program. We also appreciate support from the AFRL Information Directorate under grant F4HBKC4162G001 and the Office of the Director of National Intelligence (ODNI) and the Intelligence Advanced Research Projects Activity (IARPA), via IAA 145483. The views and conclusions contained herein are those of the authors and should not be interpreted as necessarily representing the official policies or endorsements, either expressed or implied, of ODNI, IARPA, AFRL, or the U.S. Government. The U.S. Government is authorized to reproduce and distribute reprints for Governmental purpose notwithstanding any copyright annotation thereon.
ZGI was also supported by the USRA Feynman Quantum Academy funded by the NAMS R\&D Student Program at NASA Ames Research Center.
ZGI, SH, JM, and ZW are also supported by NASA Academic Mission Services, Contract No. NNA16BD14C.
We acknowledge insightful discussions with Davide Venturelli
and also with Riccardo  Mengoni, particularly preliminary work mapping spanning tree problems to QUBO \cite{wang2020}.

\bibliography{qc,spanningTrees}

\appendix

\input{sections/level_mapping}
\input{sections/graphs_and_weights}
\input{sections/embedding_stats}

\clearpage
\section{Details on unsolved instances}
\label{appendix:unsolved_cases}
As a special case of the improvement in TTS,
we find that for certain problems, the no-pause annealing failed to find a solution even after 50K reads, while the annealing with an appropriate pause was able to find one. 
In particular, out of the 45 instances tested, the no-pause annealing failed to solve 7 of them. Of those 7, there are 3 which remained unsolved by any of the pause runs (we are considering a total of 10 pause runs, resulting from 2 pause locations $s_p = \{0.3, 0.32\}$ and 5 pause durations $t_p = \{0.25, 0.5, 0.75, 1, 2\}~\mu$s), while the other 4 were solved by most or all of them: 2 were solved by 10 out of the 10 pause runs, 1 solved by 9 pause runs, and the other one solved by 8 pause runs. There are also 2 other instances that were solved by the no pause runs but that, respectively, 1 and 3 of the pause runs could not solve (but all the rest could). There are no instances that were solved by the no pause runs but weren't solved by the pause runs. Of the 10 pause runs, the worst one cannot solve 6 of the 45 instances (making it better than the no pause in that metric). The second worst cannot solve 5, there are 2 that cannot solve 4, and the other 6 cannot solve 3.

\section{Supporting instances showcasing shift of optimal pause location, improvements with partial gauges and the effect of pause on success probability }
\label{appendix:more}
\begin{figure}[h!]
  \includegraphics[width=\linewidth]{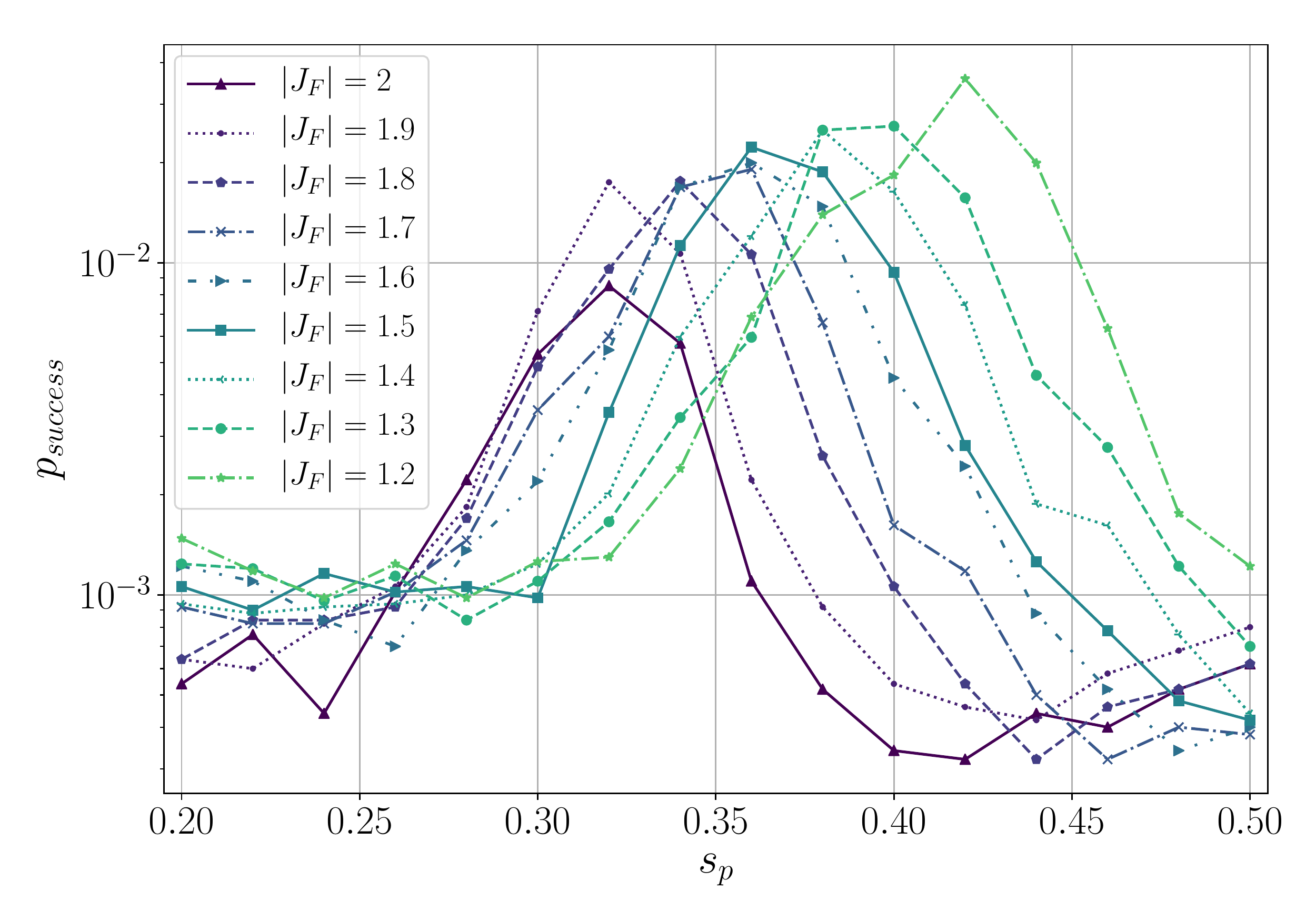}
  \caption{{\bf Optimal pause location shift for an instance with $\Delta=3$} 
  Probability of success versus the annealing pause location for the n=5, m5ver5 [m=5, $K_{1,4}+e$, g6: DiS], weight set 14 instance using embedding number 20 with $\Delta$=3, 1 $\mu$s anneal, 100 $\mu$s pause, 50K reads and 0 partial gauges. Pause location ranging from 0.2 to 0.5 and $J_{ferro}$ varied from -1.2 to -2.0. The peak in $p_{success}$ shifts from $s_p$=0.42 at $J_{ferro}$=-1.2; to $s_p$=0.36 for $J_{ferro}$=-1.5; and $s_p$=0.32 for $J_{ferro}$=-2.0. Note that $\Delta$=3 is the minimum delta that can be used to obtain a minimum weighted spanning tree.
}
  \label{fig:delta_3}
\end{figure}

In Fig.~\ref{fig:delta_3}, we show the shift of optimal pause location with $|J_F|$ for a problem instance for bounded degree $\Delta=3$.

\begin{figure*}
  \includegraphics[width=0.95\linewidth]{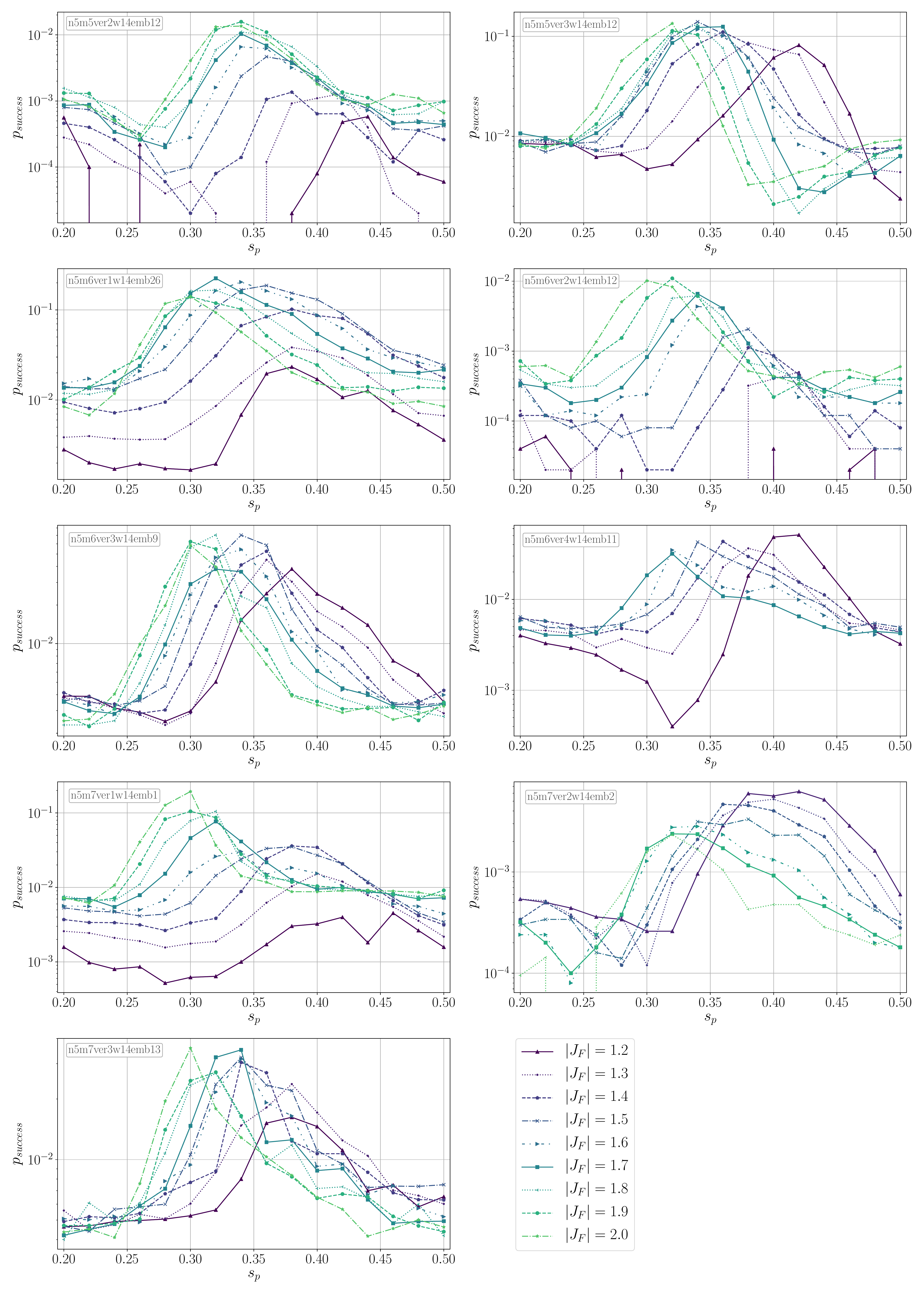}
    \caption{{\bf Shift of optimal pause location with $|J_F|$ (for multiple instances).}
    Shifting of optimal pause location with $|J_F|$ for multiple instances with a $100~\mu$s pause. Note that the scale of the y axis is different across instances.
 }
  \label{fig:more_instances}
\end{figure*}

In Fig.~\ref{fig:more_instances}, we show a few more instances from the instance ensemble for $\Delta=2$, $n=5$. 

\begin{figure}
\centering
\begin{minipage}[t]{.5\textwidth}
  \centering
  \includegraphics[width=.95\linewidth]{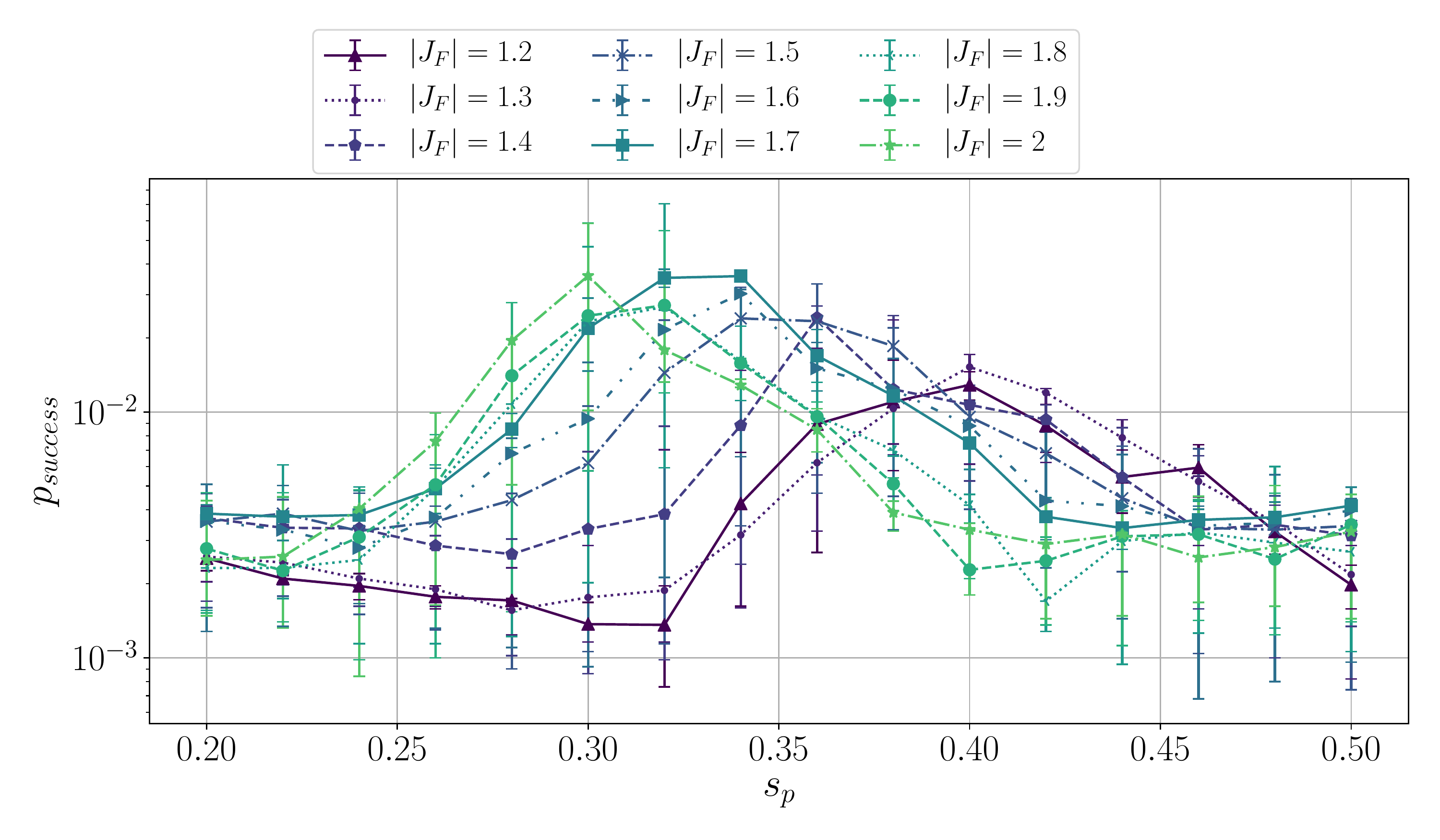}
  \captionof{figure}{{\bf Shift of optimal pause location with $|J_F|$ (ensemble).}
  Probability of success for an ensemble of $9$ instances of $n=5$ with a pause duration $t_p=100~\mu$s, and $t_a=1\mu$s.}
  \label{fig:ensemble_shift_new}
\end{minipage}
\begin{minipage}[t]{.5\textwidth}
  \centering
  \includegraphics[width=.80\linewidth]{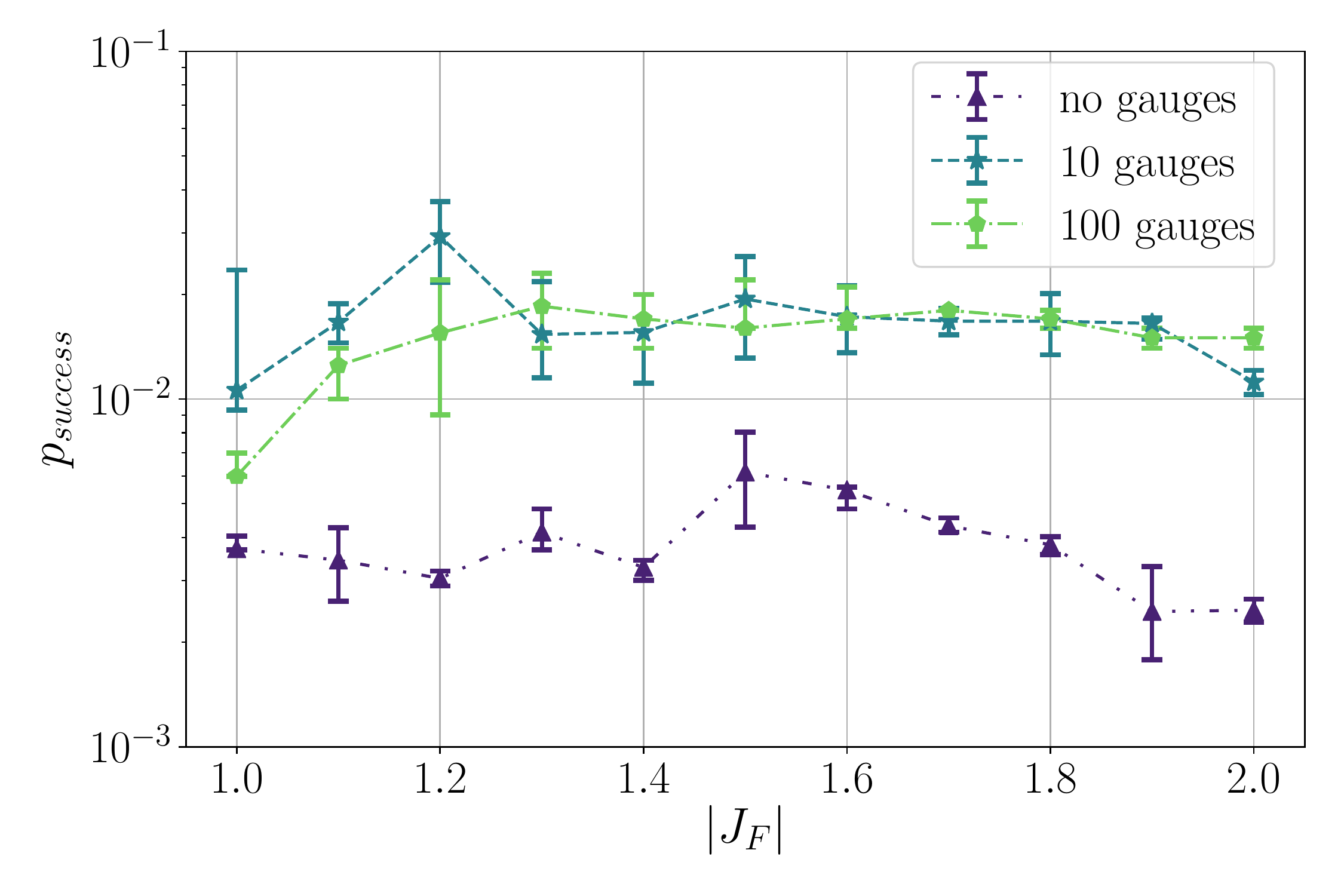}
  \captionof{figure}{{\bf Improvement of probability of success with partial gauges.} Effect of partial gauges on the probability of success for 10 of $n=4$ instances with varying $|J_F|$ and a no pause schedule.}
  \label{fig:n4_gauges}
\end{minipage}
\end{figure}

Figure~\ref{fig:ensemble_shift_new} illustrates the clear shifting of the optimal pause location for an instance ensemble over all $|J_F|$ values we examined (in range $[1.2, 2]$). For figure clarity, pausing results for just three values of $|J_F|$ are shown earlier in the bottom panel of Fig.~\ref{fig:demo_shift} and discussed in Section~\ref{sec:shift}

As discussed in Section~\ref{sec:partial_gauges}, the improvement in $p_{success}$ will saturate as one increases the number of gauges applied.  Fig.~\ref{fig:n4_gauges} shows for an $n=4$ ensemble, applying as few as 10 gauges yields similar $p_{success}$ to 100 gauges. 
As detailed in Section~\ref{sec:partial_gauges}, these results indicate that with 10 gauges we are already likely to encounter one or more positive outliers, leading to the large improvement in $p_{success}$.  As the number of gauges further increases, the effect is not as dramatic, indicating the spread in gauge quality approaches the intrinsic distribution.

\begin{figure}[h]
  \includegraphics[width=0.85\linewidth]{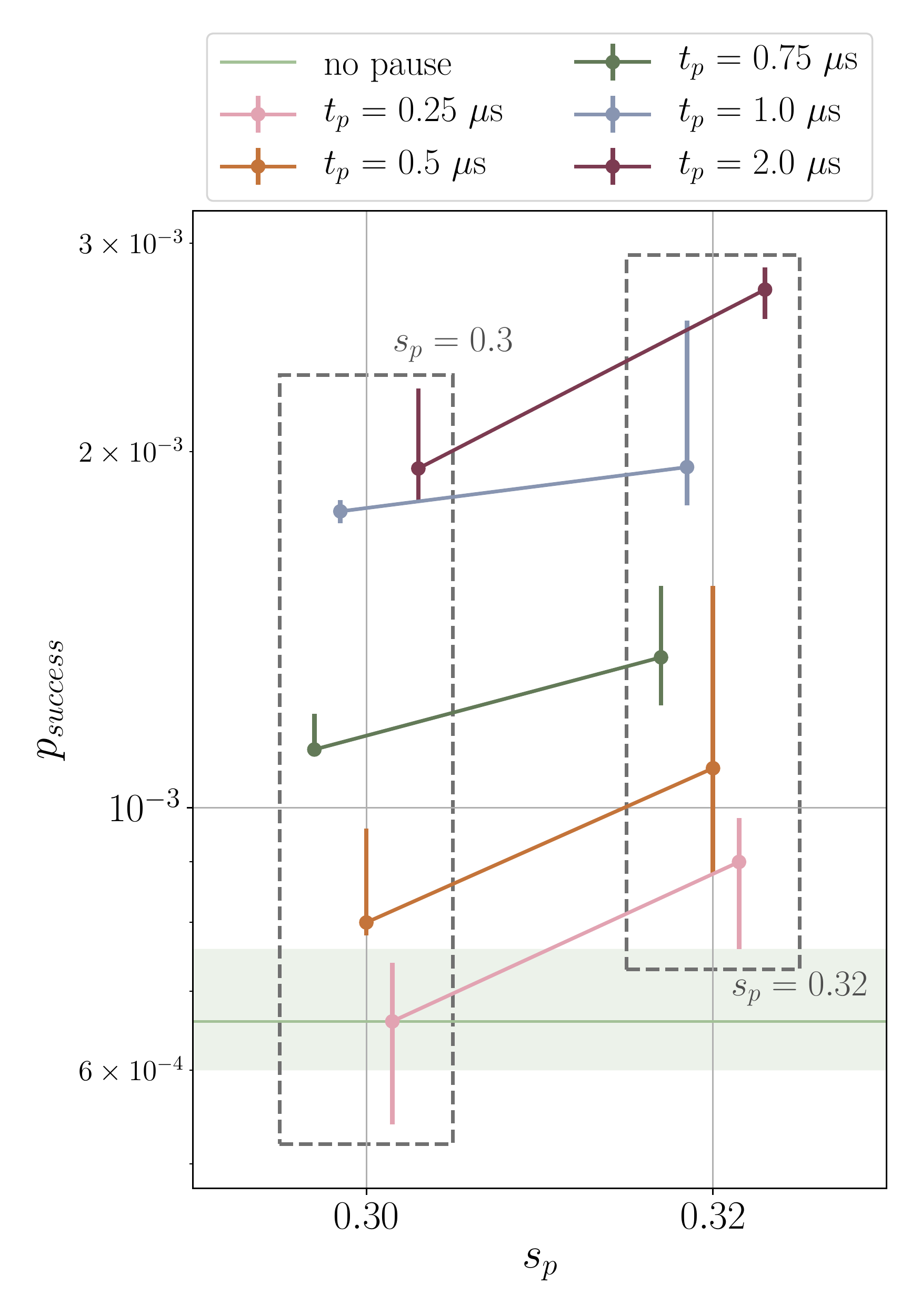}
  \caption{{\bf Effect of pause duration on success probability.}
  Success probability corresponding to TTS shown in Figure~\ref{fig:tts_improv} in Section~\ref{sec:improve_with_pause}.
With pause duration of $\{0.25, 0.5, 0.75, 1, 2\}~\mu$s, and $|J_F|$=1.8, the success probability for an ensemble of 45 instances is shown for pause locations $s_p=0.3$ and $s_p=0.32$ (which we found to be optimal during initial sweep). The reference (horizontal line and band for median and 35 and 65 percentiles, respectively) is the no-pausing case with parameters optimal for TTS: $t_a=1\mu$s, and $|J_F|$=1.6. Data points show the median, with error bars at the $35^{th}$ and $65^{th}$ percentiles, after performing $10^5$ bootstraps over the set of instances.}
  \label{fig:pause_duration}
\end{figure}

Figure~\ref{fig:tts_improv} in Section~\ref{sec:improve_with_pause} contains results for TTS for an ensemble in the narrowed parameter range discussed in this section. The corresponding $p_{success}$ are shown in  Fig.~\ref{fig:pause_duration}.

\newpage

\end{document}

%% file: header.tex
\usepackage{graphicx}
\usepackage{caption}
\usepackage{amsmath}
\usepackage{amssymb}
\usepackage{mathtools}
\usepackage{dsfont}
\usepackage{bm}
\usepackage{color}
\usepackage{appendix}

\usepackage{enumitem}
\usepackage{amsthm}

\usepackage[usenames,dvipsnames]{xcolor}
\usepackage{multirow} 

\newtheorem{theorem}{Theorem}
\numberwithin{theorem}{subsection}
\newtheorem{definition}[theorem]{Definition}

\newcommand{\bqa}{\begin{eqnarray}}
\newcommand{\eqa}{\end{eqnarray}}
\newcommand{\beq}{\begin{equation}}
\newcommand{\eeq}{\end{equation}}

\newcommand{\mst}{BD-MST }

%% file: sections/level_mapping.tex
\newpage
\section{Problem mapping: `level-based'}
\label{appendix:mapping}

Consider a graph $G=(V,E)$ with weights $w(E)$ for each edge, from which we wish to obtain a minimal weighted spanning tree with maximum degree $\Delta$, i.e. find its BD-MST. This will involve minimizing the sum of the weights of the tree edges, represented by the cost function

\begin{align}
C_0 = \sum_{p,v} w_{pv}x_{p,v},
\label{eq:cost}
\end{align}
which we explain below. Several constraints will also be imposed to ensure that the graph is in fact a spanning tree and its degree is bounded by $\Delta$.

A root for the tree will be picked randomly or based on problem structure---generally, picking a high-degree vertex as the root will result in lower resource costs---and assigned to level $1$. Its children will be at level $2$, their children at level $3$ and so on, leading to the `level-based' designation.

The variables $x_{p,v}$ appearing in Eq.~\eqref{eq:cost} represent the parent-child relationships in the tree; $x_{p,v}=1$ if $p$ is the (adjacent) parent of $v$ (and $0$ if not). The indices $p,v$ range over $p=1,\dots,n$ and $v=2,\dots,n$, restricted to (intersected with) pairs $(p,v)$ or $(v,p)$ that occur in $E$. Thus there are $2$ variables for every edge not containing the root, and one for every root edge, giving $2m-d_r$ total $x_{p,v}$ variables, with $m$ being the number of edges in $E$ and $d_r$ the degree of the root.

Since our problem needs to be in QUBO form, the constraints will be expressed as penalty terms. The first penalty term enforces that every node (except the root) has exactly one parent,
\begin{align}
 C_{pen1} = \sum_{v\in\{2,\dots,n\}}
\left(\sum_{p:(pv)\in E} x_{p,v}-1\right)^2\;.
\end{align}
The number of terms in the sum is $2m-d_r$, i.e. equal to the number of variables $x_{p,v}$.

The second penalty term enforces that each vertex exists at exactly one level in the tree,
\begin{align} 
C_{pen2} = \sum_{v\in\{2,\dots,n\}} 
\left(\sum_{\ell=2}^n y_{v,\ell}-1\right)^2\;.
\end{align}

It introduces the $y_{v,\ell}$ variables, with $y_{v,\ell}=1$ if $v$ is at depth $\ell$ of the tree, $v=2,\dots,n$, $\ell=2,\dots,n$. There are $(n-1)^2$ such variables. However, since the number of variables will eventually determine how many logical qubits the problem requires, it is in our interest to reduce it as much as possible. By picking the root smartly the range of $\ell$ can be reduced. We also carry out the following pre-processing: taking the original graph $G=(V,E)$, the distance from each node to the one we have selected as the tree root is calculated. Given that it is impossible for a node to be at a level smaller than its distance to the root, we can avoid generating any $y_{v,\ell}$ for which that is the case, further bringing down the total number of $y_{v,\ell}$ variables.

The third penalty term enforces that the tree has degree at most $\Delta$,
\begin{align}
C_{pen3} = &\sum_{p=2}^v \left(\sum_{v:(pv)\in E} x_{p,v}-\sum_{j=1}^{\Delta-1}z_{p,j}\right)^2 \\
+ &(\sum_{v:(1v)\in E} x_{1,v} - \sum_{j=1}^{\Delta}z_{1,j})^2.   
\end{align}
It is separated into two terms to account for the fact that the root can have up to $\Delta$ children, while all other nodes cannot have more than $(\Delta - 1)$, since they have a parent. 
To enforce the inequality $\sum_{v:(pv)\in E} x_{p,v} \le \Delta-1$, integer variable $z_p\in [0,\Delta-1]$ is introduced as slack variable, and the inequality is enforced as equality $\sum_{v:(pv)\in E}  x_{p,v} = z_p$.  The integer variable is further encoded into binary variables $z_{p,j}$.  In general, various encoding methods can be applied to encode an integer into binaries, including binary, unary, and one-hot encodings.  While binary encoding is most efficient for integers of value power of two, We use unary encoding here, which can be applied straightforward to arbitrary value of $\Delta$.

The fourth and final penalty term enforces that the tree encoding is consistent, i.e., that if $p$ is the parent of $v$ then its level is one less than $v$'s, 
\begin{align}
  C_{pen4} = &\sum_{p,v} 
\sum_{\ell=3}^n x_{p,v} y_{v,\ell}(1-y_{p,\ell-1}) \\ 
+ &\sum_{v=2}^{d_r} x_{1,v}(1-y_{v,2})
+ \sum_{v=2}^{d_r} y_{v,2}(1-x_{1,v})\;,
\end{align}
where the last two sums handle the edges connected to the root and their terms are quadratic, while the first sum deals with the remaining edges and produces cubic terms of the form $x_{p,v} y_{v,\ell}(1-y_{p,\ell-1})$. While the original number of cubic terms would be $$(2m-2d_r)*(n-2),$$ 
thanks to the preprocessing of the $y_{v,\ell}$ variables this number is reduced. 
Because cubic terms cannot be directly encoded in D-Wave, we introduce an ancilla variable
$a_{p,v,\ell}$ to encode  $x_{p,v} y_{v,\ell}$,
and accordingly a penalty function $f(x,y,a)=3a+xy-2ax-2ay$ is added to raise a penalty if $a=xy$ is violated.
The term $x_{p,v} y_{v,\ell}(1-y_{p,\ell-1})$ then can be replaced by quadratic terms

\begin{align} 4a - ay_{p,\ell-1}
+ x_{p,v} y_{v,\ell} - 2 ax_{p,v}- 2a y_{v,\ell}\;.
\end{align} 

The total number of variables (and hence, logical qubits) without preprocessing is at most:
\begin{align*}
  2m-d_r  + (n-1)^2 + n(\Delta-1)+1 + (2m-2d_r)(n-2) \\
  \simeq 2mn + n^2  
\end{align*}

This would mean, for instance, that the complete graph $K_5$ with $\Delta =3$ would require between $86$ and $100$ logical qubits (depending on $d_r$). With pre-processing, we are able to bring this number down to $74$.

Finally, we can write the overall objective function as
\begin{align} 
\label{eq:C_tot}
 C = C_0 + A(C_{pen1}+C_{pen2}+C_{pen3}+C_{pen4})\;,
 \end{align} 
and accordingly the cost Hamiltonian $H_C$.
In Eq.~\eqref{eq:C_tot} we have defined the minimum penalty weight to be the maximum edge weight
\begin{align} 
A=w_{max} =\max_{(uv)\in E} w_{uv}\;.
\end{align} 

In Ref.~\cite{wang2020} we provide proof that setting $A=w_{max} + \epsilon$ with any positive $\epsilon$ suffices to guarantee $C$ is minimized by bounded-degree spanning trees that are optimal for $C_0$ and correctly encoded.
In our runs, for convenience, we set $\epsilon=0$, which could in principle have led to an invalid bit string also minimizing $C$.
The solutions returned from the quantum annealer were each checked for optimality and correct encoding.
Though increasing $A$ by any amount would guarantee that the optimal cost of a solution implies its correct encoding, in practice we observed that this was still the case despite having set $\epsilon=0$. We provide 
details and further discussion in~\cite{wang2020}.

%% file: sections/graphs_and_weights.tex
\section{Approximation complexity for BD-MST Problems}
\label{sec:complexity}
Finding a degree-bounded spanning tree of cost at most~$r$ times the optimum remains NP-hard for any $r\geq 1$ \cite{ravi1993many}. 
Hence, 
approximation algorithms are often designed to return a low-weight spanning tree with 
the vertex degree bound $\Delta$ slightly relaxed.
In \cite{furer1992approximating} a polynomial time algorithm is given for the unweighted problem which returns a spanning tree 
of degree at most $\Delta^*+1$, where $\Delta^*$ is the minimal
$\Delta$ for which such a 
spanning tree exists. 
For the weighted case,
Ref.~\cite{goemans2006minimum} shows a polynomial time algorithm 
that returns a spanning tree with vertex degree at most $\Delta+2$ -- subsequently improved to $\Delta+1$ in~\cite{singh2007approximating} -- and
cost at most $OPT$, where $OPT$ is the 
optimal spanning tree weight 
under 
the desired bound~$\Delta$.
Alternatively, heuristics exist which return valid $\Delta$-bounded spanning trees, but with suboptimal cost that may be difficult to quantify generally.
A wide variety of 
approaches have been developed for this problem \cite{konemann2000matter,zahrani2008local, khuller1996low, jothi2009degree,bui2006ant,bui2011improved}, including specific approximations for various special cases (e.g., geometric weights); see 
\cite{krishnamoorthy2001comparison} for an
overview. 

\section{BD-MST Problem instances}
\label{app:instances}

All connected graphs of $n=5$, with $m=|E|$ ranging from 4 to 10 are considered where an \mst with $\Delta = 2$ exists. The edges in these graphs is provide in Table \ref{tab:n5graph}. Additionally, the graph labeled 'm5ver5' is included to demonstrate the \mst with $\Delta \geq 3$.
For each graph, problem instances were generated by assigning a set of weights by sampling from one of the lists of weights appearing in Table \ref{tab:weights}. The first $m$ weights in each weight list were used to define an instance.

\begin{table*}
\parbox[h!]{0.6\textwidth}{
\centering
\begin{tabular}{| c | c | c | c |}
\hline
 m & label & Graph6 Name & edges \\ 
 \hline
 4 & m4ver1 & DhC & (1,2), (2,3), (3,4), (4,5) \\  
 5 & m5ver1 & Dhc & (1,2), (2,3), (3,4), (4,5), (1,5) \\
 5 & m5ver2 & DiK & (1,2), (2,3), (2,5), (3,4), (4,5) \\
 5 & m5ver3 & DjC & (1,2), (2,3), (2,4), (3,4), (4,5) \\
 5 & m5ver5 & DiS & (1, 2),(1, 3), (1,4),(1,5),(4,5) \\
 5 & m5ver6 & DKs & (1, 2), (2, 3), (3, 4), (4, 5), (3, 5) \\
 6 & m6ver1 & DyK & (1,2), (1,5), (2,5), (2,3), (3,4), (4,5) \\
 6 & m6ver2 & DjS & (1,2), (2,3), (2,4), (2,5), (3,4), (4,5) \\
 6 & m6ver3 & DjK & (1,2), (2,3), (2,5), (3,5), (3,4), (4,5) \\
 6 & m6ver4 & D\{K & (1,2), (1,5), (1,3), (2,3), (3,4), (4,5) \\
 6 & m6ver5 & D\{c & (1, 2), (1, 3), (2, 3), (3, 5), (3, 4), (4, 5) \\
 6 & m6ver6 & D]o & (1, 2), (2, 3), (3, 4), (1, 4), (2, 5), (4, 5) \\
 7 & m7ver1 & D$|$S & (1,2), (1,5), (1,4), (2,5), (2,3), (3,4), (4,5)\\
 7 & m7ver2 & DzW & (1,2), (1,5), (2,5), (2,3), (2,4), (3,5), (4,5)\\
 7 & m7ver3 & D$|$c & (1,2), (1,3), (1,4), (1,5), (2,3), (3,4), (4,5)\\
 7 & m7ver4 & D$\sim$C & (1,2), (1,3), (1,4), (2,4), (2,3), (3,4), (4,5)\\
 7 & m7ver5 & D]w & (1, 2), (2, 3), (3, 4), (1, 4), (4, 5), (2, 5), (3, 5)\\
 7 & m7ver6 & Dh\{ & (1, 2), (1, 3), (1, 4), (1, 5), (2, 3), (3, 4), (4, 5)\\
 8 & m8ver1 & D\}k & (1,2), (1,5), (1,3), (1,4), (2,5), (2,3), (3,4), (4,5) \\
 8 & m8ver2 & Dz[ & (1,2), (1,5), (2,5), (2,3), (2,4), (3,4), (3,5), (4,5)\\
 9 & m9ver1 & D$\sim$k & 
    \begin{tabular}{@{}c@{}}(1, 2), (2, 3), (4, 5), (1, 5), (1, 4), \\ (1, 3), (2, 5), (2, 4), (3, 5)\end{tabular} \\
 10 & m10ver1 & D$\sim$\{ & 
    \begin{tabular}{@{}c@{}}(1, 2), (2, 3), (3, 4), (4, 5), (1, 5), \\ (1, 4), (1, 3), (2, 5), (2, 4), (3, 5)\end{tabular} \\
 \hline
\end{tabular}
\caption{$n=5$ graphs}
\label{tab:n5graph}
}
\hfill
\parbox[h!]{.32\textwidth}{
\centering
\begin{tabular}{| c | c |}
\hline
 label & weight list \\ 
 \hline
  w2 & [1, 2, 1, 2, 1, 2, 1, 2, 1, 2] \\  
  w3 & [1, 1, 2, 1, 1, 2, 1, 1, 2, 1] \\
  w4 & [1, 1, 2, 2, 1, 1, 2, 2, 1, 1] \\
  w5 & [1, 4, 1, 4, 1, 4, 1, 4, 1, 4] \\
  w6 & [1, 3, 6, 1, 3, 6, 1, 3, 6, 1] \\
  w7 & [1, 7, 1, 7, 1, 7, 1, 7, 1, 7] \\
  w8 & [3, 2, 1, 3, 2, 1, 3, 2, 1, 3] \\
  w9 & [4, 3, 2, 1, 4, 3, 2, 1, 4, 3]\\
  w10 & [5, 4, 3, 2, 1, 5, 4, 3, 2, 1]\\
  w11 & [6, 5, 4, 3, 2, 1, 6, 5, 4, 3]\\
  w12 & [7, 6, 5, 4, 3, 2, 1, 7, 6, 5]\\
  w13 & [1, 1, 3, 4, 2, 1, 2, 3, 4, 2] \\
  w14 & [3, 2, 1, 1, 1, 1, 2, 4, 2, 2]\\
  w15 & [2, 1, 2, 1, 4, 1, 1, 3, 3, 2] \\
  w16 & [4, 3, 3, 4, 3, 3, 4, 3, 4 ] \\
  w17 & [3, 4, 7, 5, 5, 5, 5] \\
  w18 & [2, 1, 4, 1, 2, 1, 2] \\
  w19 & [4, 6, 4, 7, 4, 7] \\
  w20 & [1, 1, 2, 3, 2, 3] \\
  w21 & [4, 5, 4, 5, 5] \\
  w22 & [2, 2, 6, 2, 4] \\
  w23 & [3, 3, 5, 2, 3, 2, 5, 2, 5] \\
  w24 & [4, 3, 2, 2] \\
  w25 & [2, 2, 6, 2, 4] \\
  w26 & [4, 3, 3, 3] \\
  w27 & [3, 4, 7, 5, 5, 5, 5] \\
  w28 & [4, 6, 4, 7, 4, 7] \\
  w29 & [6, 4, 2, 2] \\
\hline
\end{tabular}
\caption{Graph weights are uniformly drawn from the above lists.}
\label{tab:weights}
}
\end{table*}

%% file: sections/embedding_stats.tex
\section{Embedding Statistics}
\label{appendix:embedding}

Table \ref{tab:chimera+pegasus} contains mapped problem size for each graph, and embedding features like number of physical qubits, size of the vertex model, etc.  Embedding statistics on a future D-Wave architecture (Pegasus) are also included in this table. For the Pegasus architecture, each qubit can couple to 15 other qubits, as opposed to the Chimera architecture that allows each qubit to connect to at most 6 additional qubits.

\begin{figure}[h]
  \includegraphics[width=\linewidth]{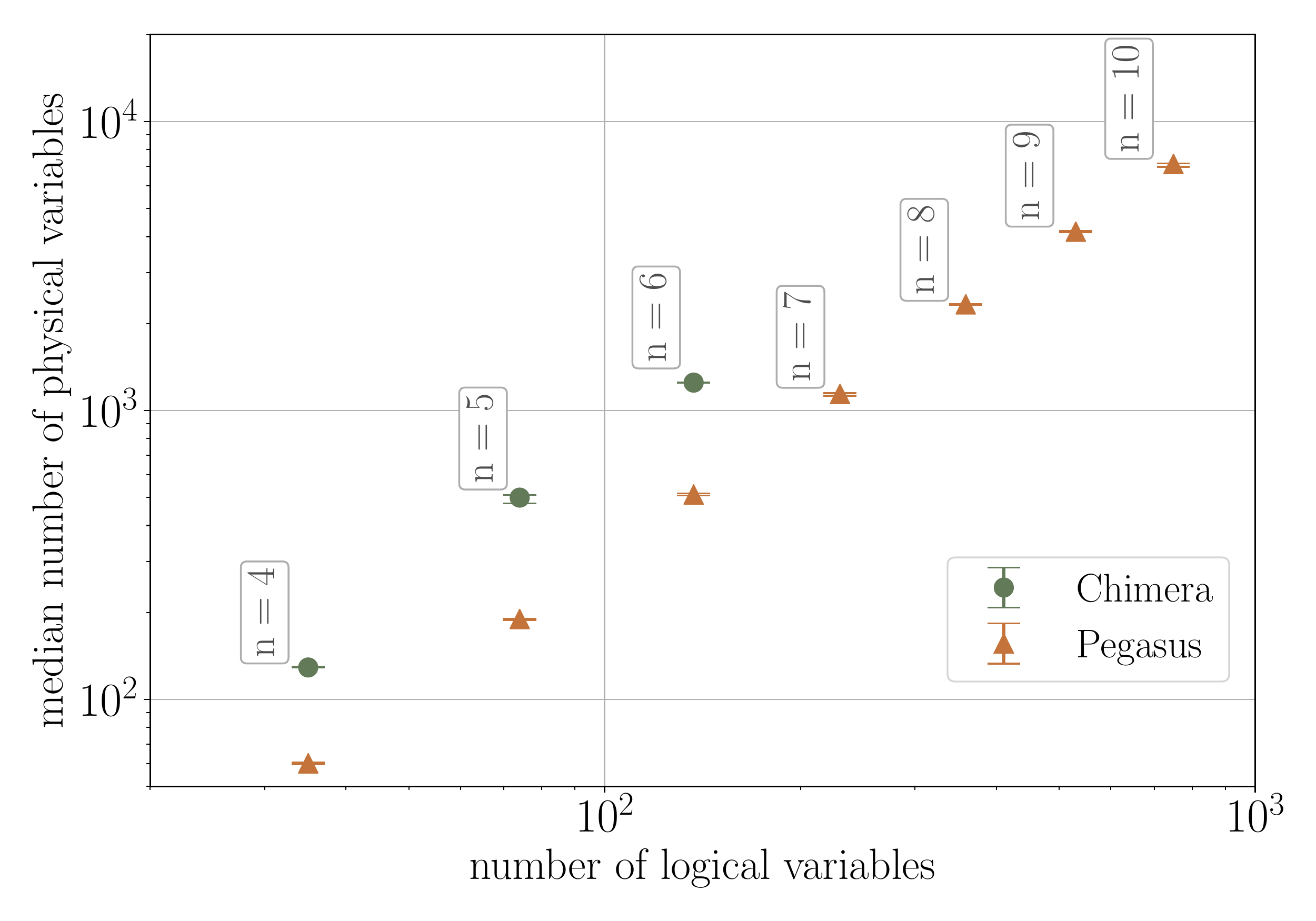}
  \caption{{\bf Embedding comparison between current and future architectures.} Embedding for the complete graphs for problem size $n=4$-$10$ with default embedding parameters and 10 to 20 instances drawn for each graph.
  Chimera embedding performed with D-Wave's SAPI2 find\_embedding routine with the D-Wave 2000Q hardware adjacency graph. Pegasus embedding performed with the Ocean minorminer find\_embedding routine. Median number of physical qubits as a function of the number of logical qubits with error bars are at the 35th and 65th percentiles after bootstrapping over the ensemble of instances.}
  \label{fig:chimera_vs_pegasus}
\end{figure}

As discussed in Section~\ref{annealing_params}, Fig.~\ref{fig:chimera_vs_pegasus} contains detailed information about the typical size of the embedded problems for different graphs, including the number of physical qubits and the size of the vertex models. Embedding statistics for a future D-Wave architecture (Pegasus) are also given.

\begin{table*}[h]
\begin{center}
\begin{tabular}{| c | c | p{2cm} | p{2cm} | p{1.5cm} | p{2cm} | p{2cm} | p{1.5cm} |}
\hline
n & m & \multicolumn{3}{c|}{Chimera Architecture} & \multicolumn{3}{c|}{Pegasus Architecture}\\ \cline{3-8}
& & \# of Logical Variables & \# of Physical Variables & Median Vertex Model Size & \# of Logical Variables & \# of Physical Variables & Median Vertex Model Size \\
\hline
 \hline
 4 & 6 & 35 & 108-150 & 3-4 & 35 & 54-71 & 1-2\\
 4 & 5 & 29 & 65-121 & - & - & - & - \\
 4 & 4 & 25 & 60-116 & - & - & - & -\\
 4 & 4 & 23 & 47-82 & - & - & - & -\\
 4 & 3 & 20 & 40-76 & - & - & - & - \\
 5 & 4 & 32 & 83-140 & 1.5-3 & - & - & - \\
 5 & 5 & 42 & 121-215 & 2-4 & - & - & - \\ 
 5 & 5 & 43 & 138-169 & 2-3 & - & - & - \\ 
 5 & 5 & 44 & 149-205 & 2-4 & - & - & - \\ 
 5 & 5 & 47 & 151-220 & 2-4 & - & - & - \\ 
 5 & 5 & 39 & 112-179 & 1.5-3 & - & - & - \\ 
 5 & 6 & 50 & 169-205 & 2-4 & - & - & - \\ 
 5 & 6 & 53 & 194-255 & 2-4 & - & - & - \\ 
 5 & 6 & 49 & 181-246 & 2-4 & - & - & - \\ 
 5 & 6 & 46 & 148-272 & 2.5-4.5 & - & - & - \\ 
 5 & 6 & 50 & 170-249 & 3-5 & - & - & - \\ 
 5 & 6 & 50 & 164-217 & 2.5-4.5 & - & - & - \\ 
 5 & 7 & 54 & 193-247 & 3-5 & - & - & - \\ 
 5 & 7 & 58 & 229-300 & 3-4.5 & - & - & - \\ 
 5 & 7 & 50 & 171-260 & 3-5 & - & - & - \\ 
 5 & 7 & 55 & 226-281 & 3-5 & - & - & - \\ 
 5 & 7 & 56 & 227-273 & 2-5 & - & - & - \\ 
 5 & 7 & 50 & 162-224 & 2.5-5 & - & - & - \\ 
 5 & 8 & 58 & 219-284 & 3-5 & - & - & - \\ 
 5 & 8 & 64 & 287-362 & 3-5 & - & - & - \\ 
 5 & 9 & 66 & 299-413 & 3.5-6 & - & - & - \\ 
 5 & 10 & 74 & 380-485 & 4-7 & 74 & 164-207 & 1-2 \\ 
 6 & 15 & 137 & 1166-1293 & 4-6 & 137 & 1166-1293 & 4-6 \\
 7 & 21 & - & - & - & 230 & 1018-1291 & 2-4 \\
 8 & 28 & - & - & - & 359 & 2046-2712 & 2-4 \\
 9 & 36 & - & - & - & 530 & 3744-4454 & 2-4 \\
 10 & 45 & - & - & - & 749 & 6024-7889 & 2-4 \\
\hline
\end{tabular}
\caption{Mapped problem size for N=4-6 using the D-Wave Chimera architecture, and problem size $N=4$-10 using the future D-Wave Pegasus architecture. For the Pegasus architecture entries, embedding was only performed for the complete graphs, which is the reason for the large number of unset entries in the last three columns. For the Chimera Architecture embedding entries, we were unable to embed graphs with $n\ge7$ using the default embedding parameters, which is the reason for the missing data entries in the last four rows of the table in columns 3-5. Lastly, we were not collecting median vertex model size statistics for some of the early $n=4$ network communication graphs that we examined early in the study, which is the reason for the missing Chimera Architecture entries for the $n=4$ graphs near the top of the table.}
\label{tab:chimera+pegasus}
\end{center}
\end{table*}

%% file: main.bbl
{\catcode`\/=\active \catcode`\.=\active \catcode`\-=\active
  \catcode`\@=\active \gdef\url{\tt\catcode`\/=\active \catcode`\.=\active
  \catcode`\-=\active \catcode`\@=\active
  \def/{\discretionary{\char`\/}{}{\char`\/}}
  \def.{\discretionary{\char`\.}{}{\char`\.}}
  \def-{\discretionary{\char`\-}{}{\char`\-}}
  \def@{\discretionary{\char`\@}{}{\char`\@}}}} \def\annote#1{}
  \def\tilde{\char126}
\begin{thebibliography}{30}%
\makeatletter
\providecommand \@ifxundefined [1]{%
 \@ifx{#1\undefined}
}%
\providecommand \@ifnum [1]{%
 \ifnum #1\expandafter \@firstoftwo
 \else \expandafter \@secondoftwo
 \fi
}%
\providecommand \@ifx [1]{%
 \ifx #1\expandafter \@firstoftwo
 \else \expandafter \@secondoftwo
 \fi
}%
\providecommand \natexlab [1]{#1}%
\providecommand \enquote  [1]{``#1''}%
\providecommand \bibnamefont  [1]{#1}%
\providecommand \bibfnamefont [1]{#1}%
\providecommand \citenamefont [1]{#1}%
\providecommand \href@noop [0]{\@secondoftwo}%
\providecommand \href [0]{\begingroup \@sanitize@url \@href}%
\providecommand \@href[1]{\@@startlink{#1}\@@href}%
\providecommand \@@href[1]{\endgroup#1\@@endlink}%
\providecommand \@sanitize@url [0]{\catcode `\\12\catcode `\$12\catcode
  `\&12\catcode `\#12\catcode `\^12\catcode `\_12\catcode `\%12\relax}%
\providecommand \@@startlink[1]{}%
\providecommand \@@endlink[0]{}%
\providecommand \url  [0]{\begingroup\@sanitize@url \@url }%
\providecommand \@url [1]{\endgroup\@href {#1}{\urlprefix }}%
\providecommand \urlprefix  [0]{URL }%
\providecommand \Eprint [0]{\href }%
\providecommand \doibase [0]{http://dx.doi.org/}%
\providecommand \selectlanguage [0]{\@gobble}%
\providecommand \bibinfo  [0]{\@secondoftwo}%
\providecommand \bibfield  [0]{\@secondoftwo}%
\providecommand \translation [1]{[#1]}%
\providecommand \BibitemOpen [0]{}%
\providecommand \bibitemStop [0]{}%
\providecommand \bibitemNoStop [0]{.\EOS\space}%
\providecommand \EOS [0]{\spacefactor3000\relax}%
\providecommand \BibitemShut  [1]{\csname bibitem#1\endcsname}%
\let\auto@bib@innerbib\@empty
\bibitem [{\citenamefont {Marshall}\ \emph {et~al.}(2019)\citenamefont
  {Marshall}, \citenamefont {Venturelli}, \citenamefont {Hen},\ and\
  \citenamefont {Rieffel}}]{Marshall19_Pausing}%
  \BibitemOpen
  \bibfield  {author} {\bibinfo {author} {\bibfnamefont {J.}~\bibnamefont
  {Marshall}}, \bibinfo {author} {\bibfnamefont {D.}~\bibnamefont
  {Venturelli}}, \bibinfo {author} {\bibfnamefont {I.}~\bibnamefont {Hen}}, \
  and\ \bibinfo {author} {\bibfnamefont {E.~G.}\ \bibnamefont {Rieffel}},\
  }\bibfield  {title} {\enquote {\bibinfo {title} {{Power of Pausing: Advancing
  Understanding of Thermalization in Experimental Quantum Annealers}},}\ }\href
  {\doibase 10.1103/PhysRevApplied.11.044083} {\bibfield  {journal} {\bibinfo
  {journal} {Phys. Rev. Applied}\ }\textbf {\bibinfo {volume} {11}},\ \bibinfo
  {pages} {044083} (\bibinfo {year} {2019})}\BibitemShut {NoStop}%
\bibitem [{\citenamefont {Passarelli}\ \emph {et~al.}(2019)\citenamefont
  {Passarelli}, \citenamefont {Cataudella},\ and\ \citenamefont
  {Lucignano}}]{p-spin-pause}%
  \BibitemOpen
  \bibfield  {author} {\bibinfo {author} {\bibfnamefont {G.}~\bibnamefont
  {Passarelli}}, \bibinfo {author} {\bibfnamefont {V.}~\bibnamefont
  {Cataudella}}, \ and\ \bibinfo {author} {\bibfnamefont {P.}~\bibnamefont
  {Lucignano}},\ }\bibfield  {title} {\enquote {\bibinfo {title} {{Improving
  quantum annealing of the ferromagnetic $p$-spin model through pausing}},}\
  }\href {\doibase 10.1103/PhysRevB.100.024302} {\bibfield  {journal} {\bibinfo
   {journal} {Phys. Rev. B}\ }\textbf {\bibinfo {volume} {100}},\ \bibinfo
  {pages} {024302} (\bibinfo {year} {2019})}\BibitemShut {NoStop}%
\bibitem [{\citenamefont {Chen}\ and\ \citenamefont
  {Lidar}(2020)}]{chen-pausing}%
  \BibitemOpen
  \bibfield  {author} {\bibinfo {author} {\bibfnamefont {H.}~\bibnamefont
  {Chen}}\ and\ \bibinfo {author} {\bibfnamefont {D.A.}\ \bibnamefont
  {Lidar}},\ }\bibfield  {title} {\enquote {\bibinfo {title} {Why and when is
  pausing beneficial in quantum annealing?}}\ }\href
  {https://arxiv.org/abs/2005.01888} {\bibfield  {journal} {\bibinfo  {journal}
  {arXiv:2005.01888}\ } (\bibinfo {year} {2020})}\BibitemShut {NoStop}%
\bibitem [{\citenamefont {Vazirani}(2013)}]{vazirani2013approximation}%
  \BibitemOpen
  \bibfield  {author} {\bibinfo {author} {\bibfnamefont {Vijay~V}\ \bibnamefont
  {Vazirani}},\ }\href {https://books.google.com/books?id=bJmqCAAAQBAJ} {\emph
  {\bibinfo {title} {Approximation algorithms}}}\ (\bibinfo  {publisher}
  {Springer Science \& Business Media},\ \bibinfo {year} {2013})\BibitemShut
  {NoStop}%
\bibitem [{\citenamefont {Cormen}\ \emph {et~al.}(2009)\citenamefont {Cormen},
  \citenamefont {Leiserson}, \citenamefont {Rivest},\ and\ \citenamefont
  {Stein}}]{cormen2009introduction}%
  \BibitemOpen
  \bibfield  {author} {\bibinfo {author} {\bibfnamefont {Thomas~H}\
  \bibnamefont {Cormen}}, \bibinfo {author} {\bibfnamefont {Charles~E}\
  \bibnamefont {Leiserson}}, \bibinfo {author} {\bibfnamefont {Ronald~L}\
  \bibnamefont {Rivest}}, \ and\ \bibinfo {author} {\bibfnamefont {Clifford}\
  \bibnamefont {Stein}},\ }\href@noop {} {\emph {\bibinfo {title} {Introduction
  to algorithms}}}\ (\bibinfo  {publisher} {MIT press},\ \bibinfo {year}
  {2009})\BibitemShut {NoStop}%
\bibitem [{\citenamefont {Garey}\ and\ \citenamefont
  {Johnson}(1979)}]{GareyJohnson}%
  \BibitemOpen
  \bibfield  {author} {\bibinfo {author} {\bibfnamefont {Michael~R.}\
  \bibnamefont {Garey}}\ and\ \bibinfo {author} {\bibfnamefont {David~S.}\
  \bibnamefont {Johnson}},\ }\href {https://dl.acm.org/doi/10.5555/574848}
  {\emph {\bibinfo {title} {Computers and Intractability: A Guide to the Theory
  of NP-Completeness}}}\ (\bibinfo  {publisher} {W. H. Freeman \& Co.},\
  \bibinfo {address} {New York, NY, USA},\ \bibinfo {year} {1979})\BibitemShut
  {NoStop}%
\bibitem [{\citenamefont {Choi}(2020)}]{Choi19}%
  \BibitemOpen
  \bibfield  {author} {\bibinfo {author} {\bibfnamefont {V.}~\bibnamefont
  {Choi}},\ }\bibfield  {title} {\enquote {\bibinfo {title} {{The effects of
  the problem Hamiltonian parameters on the minimum spectral gap in adiabatic
  quantum optimization}},}\ }\href {\doibase
  https://doi.org/10.1007/s11128-020-2582-1} {\bibfield  {journal} {\bibinfo
  {journal} {Quant. Inf. Process.}\ }\textbf {\bibinfo {volume} {19}},\
  \bibinfo {pages} {90} (\bibinfo {year} {2020})}\BibitemShut {NoStop}%
\bibitem [{\citenamefont {Rieffel}\ \emph {et~al.}(2015)\citenamefont
  {Rieffel}, \citenamefont {Venturelli}, \citenamefont {O'Gorman},
  \citenamefont {Do}, \citenamefont {Prystay},\ and\ \citenamefont
  {Smelyanskiy}}]{Rieffel14CaseStudy}%
  \BibitemOpen
  \bibfield  {author} {\bibinfo {author} {\bibfnamefont {Eleanor~G.}\
  \bibnamefont {Rieffel}}, \bibinfo {author} {\bibfnamefont {Davide}\
  \bibnamefont {Venturelli}}, \bibinfo {author} {\bibfnamefont {Bryan}\
  \bibnamefont {O'Gorman}}, \bibinfo {author} {\bibfnamefont {Minh~B.}\
  \bibnamefont {Do}}, \bibinfo {author} {\bibfnamefont {Elicia~M.}\
  \bibnamefont {Prystay}}, \ and\ \bibinfo {author} {\bibfnamefont {Vadim~N.}\
  \bibnamefont {Smelyanskiy}},\ }\bibfield  {title} {\enquote {\bibinfo {title}
  {A case study in programming a quantum annealer for hard operational planning
  problems},}\ }\href {https://dl.acm.org/doi/10.1007/s11128-014-0892-x}
  {\bibfield  {journal} {\bibinfo  {journal} {Quantum Information Processing}\
  }\textbf {\bibinfo {volume} {14}},\ \bibinfo {pages} {1--36} (\bibinfo {year}
  {2015})}\BibitemShut {NoStop}%
\bibitem [{\citenamefont {Lucas}(2014)}]{STMapping}%
  \BibitemOpen
  \bibfield  {author} {\bibinfo {author} {\bibfnamefont {Andrew}\ \bibnamefont
  {Lucas}},\ }\bibfield  {title} {\enquote {\bibinfo {title} {Ising
  formulations of many {NP} problems},}\ }\href {\doibase
  10.3389/fphy.2014.00005} {\bibfield  {journal} {\bibinfo  {journal}
  {Frontiers in Physics}\ }\textbf {\bibinfo {volume} {2}},\ \bibinfo {pages}
  {5} (\bibinfo {year} {2014})}\BibitemShut {NoStop}%
\bibitem [{\citenamefont {Choi}(2008)}]{Choi}%
  \BibitemOpen
  \bibfield  {author} {\bibinfo {author} {\bibfnamefont {V.}~\bibnamefont
  {Choi}},\ }\bibfield  {title} {\enquote {\bibinfo {title} {Minor-embedding in
  adiabatic quantum computation: I. the parameter setting problem},}\ }\href
  {\doibase https://doi.org/10.1007/s11128-008-0082-9} {\bibfield  {journal}
  {\bibinfo  {journal} {Quantum Inf. Process.}\ }\textbf {\bibinfo {volume}
  {7}},\ \bibinfo {pages} {193} (\bibinfo {year} {2008})}\BibitemShut {NoStop}%
\bibitem [{\citenamefont {Fang}\ and\ \citenamefont
  {Warburton}(2019)}]{Warburton}%
  \BibitemOpen
  \bibfield  {author} {\bibinfo {author} {\bibfnamefont {Yan-Long}\
  \bibnamefont {Fang}}\ and\ \bibinfo {author} {\bibfnamefont {P.}~\bibnamefont
  {Warburton}},\ }\bibfield  {title} {\enquote {\bibinfo {title} {Minimizing
  minor embedding energy: an application in quantum annealing},}\ }\href
  {https://arxiv.org/abs/1905.03291} {\bibfield  {journal} {\bibinfo  {journal}
  {arXiv:1905.03291}\ } (\bibinfo {year} {2019})}\BibitemShut {NoStop}%
\bibitem [{\citenamefont {Venturelli}\ \emph {et~al.}(2015)\citenamefont
  {Venturelli}, \citenamefont {Mandrà}, \citenamefont {Knysh}, \citenamefont
  {O’Gorman}, \citenamefont {Biswas},\ and\ \citenamefont
  {Smelyanskiy}}]{Venturelli15}%
  \BibitemOpen
  \bibfield  {author} {\bibinfo {author} {\bibfnamefont {D.}~\bibnamefont
  {Venturelli}}, \bibinfo {author} {\bibfnamefont {S.}~\bibnamefont {Mandrà}},
  \bibinfo {author} {\bibfnamefont {S.}~\bibnamefont {Knysh}}, \bibinfo
  {author} {\bibfnamefont {B.}~\bibnamefont {O’Gorman}}, \bibinfo {author}
  {\bibfnamefont {R.}~\bibnamefont {Biswas}}, \ and\ \bibinfo {author}
  {\bibfnamefont {V.}~\bibnamefont {Smelyanskiy}},\ }\bibfield  {title}
  {\enquote {\bibinfo {title} {{Quantum optimization of fully-connected spin
  glasses}},}\ }\href {\doibase 10.1103/PhysRevX.5.031040} {\bibfield
  {journal} {\bibinfo  {journal} {Phys. Rev. X}\ }\textbf {\bibinfo {volume}
  {5}},\ \bibinfo {pages} {031040} (\bibinfo {year} {2015})}\BibitemShut
  {NoStop}%
\bibitem [{\citenamefont {Wang}\ \emph {et~al.}(2020)\citenamefont {Wang},
  \citenamefont {Adnane}, \citenamefont {O’Gorman}, \citenamefont {Hadfield},
  \citenamefont {Mengoni}, \citenamefont {Venurelli}, \citenamefont
  {Izguierdo},\ and\ \citenamefont {Rieffel}}]{wang2020}%
  \BibitemOpen
  \bibfield  {author} {\bibinfo {author} {\bibfnamefont {Zhihui}\ \bibnamefont
  {Wang}}, \bibinfo {author} {\bibfnamefont {Mostafa}\ \bibnamefont {Adnane}},
  \bibinfo {author} {\bibfnamefont {Bryan}\ \bibnamefont {O’Gorman}},
  \bibinfo {author} {\bibfnamefont {Stuart}\ \bibnamefont {Hadfield}}, \bibinfo
  {author} {\bibfnamefont {Riccardo}\ \bibnamefont {Mengoni}}, \bibinfo
  {author} {\bibfnamefont {Davide}\ \bibnamefont {Venurelli}}, \bibinfo
  {author} {\bibfnamefont {Zoe~Gonzalez}\ \bibnamefont {Izguierdo}}, \ and\
  \bibinfo {author} {\bibfnamefont {Eleanor}\ \bibnamefont {Rieffel}},\
  }\href@noop {} {\enquote {\bibinfo {title} {Mapping spanning graph problems
  to quantum heuristics: techniques for handling global connectivity},}\
  }\bibinfo {howpublished} {In preparation.} (\bibinfo {year}
  {2020})\BibitemShut {NoStop}%
\bibitem [{DWa(2019)}]{DWave}%
  \BibitemOpen
  \href@noop {} {\enquote {\bibinfo {title} {{QPU Properties: D-Wave 2000Q
  System at NASA Ames}},}\ }\bibinfo {howpublished} {D-Wave User Manual
  09-1151A-D} (\bibinfo {year} {2019})\BibitemShut {NoStop}%
\bibitem [{\citenamefont {Boixo}\ \emph {et~al.}(2014)\citenamefont {Boixo},
  \citenamefont {R{\o}nnow}, \citenamefont {Isakov}, \citenamefont {Wang},
  \citenamefont {Wecker}, \citenamefont {Lidar}, \citenamefont {Martinis},\
  and\ \citenamefont {Troyer}}]{Boixo2014evidence}%
  \BibitemOpen
  \bibfield  {author} {\bibinfo {author} {\bibfnamefont {Sergio}\ \bibnamefont
  {Boixo}}, \bibinfo {author} {\bibfnamefont {Troels~F}\ \bibnamefont
  {R{\o}nnow}}, \bibinfo {author} {\bibfnamefont {Sergei~V}\ \bibnamefont
  {Isakov}}, \bibinfo {author} {\bibfnamefont {Zhihui}\ \bibnamefont {Wang}},
  \bibinfo {author} {\bibfnamefont {David}\ \bibnamefont {Wecker}}, \bibinfo
  {author} {\bibfnamefont {Daniel~A}\ \bibnamefont {Lidar}}, \bibinfo {author}
  {\bibfnamefont {John~M}\ \bibnamefont {Martinis}}, \ and\ \bibinfo {author}
  {\bibfnamefont {Matthias}\ \bibnamefont {Troyer}},\ }\bibfield  {title}
  {\enquote {\bibinfo {title} {Evidence for quantum annealing with more than
  one hundred qubits},}\ }\href {\doibase 10.1038/nphys2900} {\bibfield
  {journal} {\bibinfo  {journal} {Nature Physics}\ }\textbf {\bibinfo {volume}
  {10}},\ \bibinfo {pages} {218--224} (\bibinfo {year} {2014})}\BibitemShut
  {NoStop}%
\bibitem [{\citenamefont {R{\o}nnow}\ \emph {et~al.}(2014)\citenamefont
  {R{\o}nnow}, \citenamefont {Wang}, \citenamefont {Job}, \citenamefont
  {Boixo}, \citenamefont {Isakov}, \citenamefont {Wecker}, \citenamefont
  {Martinis}, \citenamefont {Lidar},\ and\ \citenamefont
  {Troyer}}]{Ronnow2014defining}%
  \BibitemOpen
  \bibfield  {author} {\bibinfo {author} {\bibfnamefont {Troels~F.}\
  \bibnamefont {R{\o}nnow}}, \bibinfo {author} {\bibfnamefont {Zhihui}\
  \bibnamefont {Wang}}, \bibinfo {author} {\bibfnamefont {Joshua}\ \bibnamefont
  {Job}}, \bibinfo {author} {\bibfnamefont {Sergio}\ \bibnamefont {Boixo}},
  \bibinfo {author} {\bibfnamefont {Sergei~V.}\ \bibnamefont {Isakov}},
  \bibinfo {author} {\bibfnamefont {David}\ \bibnamefont {Wecker}}, \bibinfo
  {author} {\bibfnamefont {John~M.}\ \bibnamefont {Martinis}}, \bibinfo
  {author} {\bibfnamefont {Daniel~A.}\ \bibnamefont {Lidar}}, \ and\ \bibinfo
  {author} {\bibfnamefont {Matthias}\ \bibnamefont {Troyer}},\ }\bibfield
  {title} {\enquote {\bibinfo {title} {Defining and detecting quantum
  speedup},}\ }\href {http://www.sciencemag.org/content/345/6195/420}
  {\bibfield  {journal} {\bibinfo  {journal} {Science}\ }\textbf {\bibinfo
  {volume} {345}},\ \bibinfo {pages} {420--424} (\bibinfo {year}
  {2014})}\BibitemShut {NoStop}%
\bibitem [{\citenamefont {Boixo}\ \emph {et~al.}(2013)\citenamefont {Boixo},
  \citenamefont {Albash}, \citenamefont {Spedalieri}, \citenamefont
  {Chancellor},\ and\ \citenamefont {Lidar}}]{Boixo2013experimental}%
  \BibitemOpen
  \bibfield  {author} {\bibinfo {author} {\bibfnamefont {Sergio}\ \bibnamefont
  {Boixo}}, \bibinfo {author} {\bibfnamefont {Tameem}\ \bibnamefont {Albash}},
  \bibinfo {author} {\bibfnamefont {Federico~M}\ \bibnamefont {Spedalieri}},
  \bibinfo {author} {\bibfnamefont {Nicholas}\ \bibnamefont {Chancellor}}, \
  and\ \bibinfo {author} {\bibfnamefont {Daniel~A}\ \bibnamefont {Lidar}},\
  }\bibfield  {title} {\enquote {\bibinfo {title} {Experimental signature of
  programmable quantum annealing},}\ }\href {\doibase 10.1038/ncomms3067}
  {\bibfield  {journal} {\bibinfo  {journal} {Nature communications}\ }\textbf
  {\bibinfo {volume} {4}},\ \bibinfo {pages} {2067} (\bibinfo {year}
  {2013})}\BibitemShut {NoStop}%
\bibitem [{\citenamefont {Kim}\ \emph {et~al.}(2019)\citenamefont {Kim},
  \citenamefont {Venturelli},\ and\ \citenamefont {Jamieson}}]{kim2019}%
  \BibitemOpen
  \bibfield  {author} {\bibinfo {author} {\bibfnamefont {M.}~\bibnamefont
  {Kim}}, \bibinfo {author} {\bibfnamefont {D.}~\bibnamefont {Venturelli}}, \
  and\ \bibinfo {author} {\bibfnamefont {K.}~\bibnamefont {Jamieson}},\
  }\bibfield  {title} {\enquote {\bibinfo {title} {{Leveraging Quantum
  Annealing for Large MIMO Processing in Centralized Radio Access Networks}},}\
  }in\ \href {\doibase 10.1145/3341302.3342072} {\emph {\bibinfo {booktitle}
  {Proceedings of the ACM Special Interest Group on Data Communication}}},\
  \bibinfo {series and number} {SIGCOMM ’19}\ (\bibinfo  {publisher}
  {Association for Computing Machinery},\ \bibinfo {address} {New York, NY,
  USA},\ \bibinfo {year} {2019})\ p.\ \bibinfo {pages} {241–255}\BibitemShut
  {NoStop}%
\bibitem [{\citenamefont {Marshall}\ \emph {et~al.}(2020)\citenamefont
  {Marshall}, \citenamefont {Di~Gioacchino},\ and\ \citenamefont
  {Rieffel}}]{marshall2020perils}%
  \BibitemOpen
  \bibfield  {author} {\bibinfo {author} {\bibfnamefont {J.}~\bibnamefont
  {Marshall}}, \bibinfo {author} {\bibfnamefont {A.}~\bibnamefont
  {Di~Gioacchino}}, \ and\ \bibinfo {author} {\bibfnamefont {E.~G.}\
  \bibnamefont {Rieffel}},\ }\bibfield  {title} {\enquote {\bibinfo {title}
  {Perils of embedding for sampling problems},}\ }\href {\doibase
  10.1103/PhysRevResearch.2.023020} {\bibfield  {journal} {\bibinfo  {journal}
  {Phys. Rev. Research}\ }\textbf {\bibinfo {volume} {2}},\ \bibinfo {pages}
  {023020} (\bibinfo {year} {2020})}\BibitemShut {NoStop}%
\bibitem [{\citenamefont {Ravi}\ \emph {et~al.}(1993)\citenamefont {Ravi},
  \citenamefont {Marathe}, \citenamefont {Ravi}, \citenamefont {Rosenkrantz},\
  and\ \citenamefont {Hunt~III}}]{ravi1993many}%
  \BibitemOpen
  \bibfield  {author} {\bibinfo {author} {\bibfnamefont {R}~\bibnamefont
  {Ravi}}, \bibinfo {author} {\bibfnamefont {Madhav~V}\ \bibnamefont
  {Marathe}}, \bibinfo {author} {\bibfnamefont {SS}~\bibnamefont {Ravi}},
  \bibinfo {author} {\bibfnamefont {Daniel~J}\ \bibnamefont {Rosenkrantz}}, \
  and\ \bibinfo {author} {\bibfnamefont {Harry~B}\ \bibnamefont {Hunt~III}},\
  }\bibfield  {title} {\enquote {\bibinfo {title} {Many birds with one stone:
  Multi-objective approximation algorithms},}\ }in\ \href {\doibase
  10.1145/167088.167209} {\emph {\bibinfo {booktitle} {Proceedings of the
  twenty-fifth annual ACM symposium on Theory of computing}}},\ \bibinfo
  {organization} {ACM}\ (\bibinfo  {publisher} {Association for Computing
  Machinery},\ \bibinfo {address} {New York, NY, USA},\ \bibinfo {year}
  {1993})\ pp.\ \bibinfo {pages} {438--447}\BibitemShut {NoStop}%
\bibitem [{\citenamefont {F{\"u}rer}\ and\ \citenamefont
  {Raghavachari}(1992)}]{furer1992approximating}%
  \BibitemOpen
  \bibfield  {author} {\bibinfo {author} {\bibfnamefont {Martin}\ \bibnamefont
  {F{\"u}rer}}\ and\ \bibinfo {author} {\bibfnamefont {Balaji}\ \bibnamefont
  {Raghavachari}},\ }\bibfield  {title} {\enquote {\bibinfo {title}
  {Approximating the minimum degree spanning tree to within one from the
  optimal degree},}\ }in\ \href {https://dl.acm.org/doi/10.5555/139404.139469}
  {\emph {\bibinfo {booktitle} {Proceedings of the third annual {ACM-SIAM}
  Symposium on Discrete Algorithms}}}\ (\bibinfo {organization} {Society for
  Industrial and Applied Mathematics},\ \bibinfo {year} {1992})\ pp.\ \bibinfo
  {pages} {317--324}\BibitemShut {NoStop}%
\bibitem [{\citenamefont {Goemans}(2006)}]{goemans2006minimum}%
  \BibitemOpen
  \bibfield  {author} {\bibinfo {author} {\bibfnamefont {Michel~X}\
  \bibnamefont {Goemans}},\ }\bibfield  {title} {\enquote {\bibinfo {title}
  {Minimum bounded degree spanning trees},}\ }in\ \href {\doibase
  10.1109/FOCS.2006.48} {\emph {\bibinfo {booktitle} {Proceedings of the 47th
  Annual IEEE Symposium on Foundations of Computer Science}}},\ \bibinfo
  {series and number} {FOCS ’06},\ \bibinfo {organization} {IEEE}\ (\bibinfo
  {publisher} {IEEE Computer Society},\ \bibinfo {address} {USA},\ \bibinfo
  {year} {2006})\ pp.\ \bibinfo {pages} {273--282}\BibitemShut {NoStop}%
\bibitem [{\citenamefont {Singh}\ and\ \citenamefont
  {Lau}(2007)}]{singh2007approximating}%
  \BibitemOpen
  \bibfield  {author} {\bibinfo {author} {\bibfnamefont {Mohit}\ \bibnamefont
  {Singh}}\ and\ \bibinfo {author} {\bibfnamefont {Lap~Chi}\ \bibnamefont
  {Lau}},\ }\bibfield  {title} {\enquote {\bibinfo {title} {Approximating
  minimum bounded degree spanning trees to within one of optimal},}\ }in\ \href
  {\doibase 10.1145/1250790.1250887} {\emph {\bibinfo {booktitle} {Proceedings
  of the thirty-ninth annual ACM symposium on Theory of computing}}},\ \bibinfo
  {organization} {ACM}\ (\bibinfo  {publisher} {Association for Computing
  Machinery},\ \bibinfo {address} {New York, NY, USA},\ \bibinfo {year}
  {2007})\ pp.\ \bibinfo {pages} {661--670}\BibitemShut {NoStop}%
\bibitem [{\citenamefont {K{\"o}nemann}\ and\ \citenamefont
  {Ravi}(2000)}]{konemann2000matter}%
  \BibitemOpen
  \bibfield  {author} {\bibinfo {author} {\bibfnamefont {Jochen}\ \bibnamefont
  {K{\"o}nemann}}\ and\ \bibinfo {author} {\bibfnamefont {R}~\bibnamefont
  {Ravi}},\ }\bibfield  {title} {\enquote {\bibinfo {title} {A matter of
  degree: Improved approximation algorithms for degree-bounded minimum spanning
  trees},}\ }in\ \href {\doibase 10.1145/335305.335371} {\emph {\bibinfo
  {booktitle} {Proceedings of the Thirty-Second Annual ACM Symposium on Theory
  of Computing}}},\ \bibinfo {organization} {ACM}\ (\bibinfo  {publisher}
  {Association for Computing Machinery},\ \bibinfo {address} {New York, NY,
  USA},\ \bibinfo {year} {2000})\ pp.\ \bibinfo {pages} {537--546}\BibitemShut
  {NoStop}%
\bibitem [{\citenamefont {Zahrani}\ \emph {et~al.}(2008)\citenamefont
  {Zahrani}, \citenamefont {Loomes}, \citenamefont {Malcolm},\ and\
  \citenamefont {Albrecht}}]{zahrani2008local}%
  \BibitemOpen
  \bibfield  {author} {\bibinfo {author} {\bibfnamefont {MS}~\bibnamefont
  {Zahrani}}, \bibinfo {author} {\bibfnamefont {Martin~J}\ \bibnamefont
  {Loomes}}, \bibinfo {author} {\bibfnamefont {JA}~\bibnamefont {Malcolm}}, \
  and\ \bibinfo {author} {\bibfnamefont {Andreas~A}\ \bibnamefont {Albrecht}},\
  }\bibfield  {title} {\enquote {\bibinfo {title} {A local search heuristic for
  bounded-degree minimum spanning trees},}\ }\href {\doibase
  10.1080/03052150802317440} {\bibfield  {journal} {\bibinfo  {journal}
  {Engineering Optimization}\ }\textbf {\bibinfo {volume} {40}},\ \bibinfo
  {pages} {1115--1135} (\bibinfo {year} {2008})}\BibitemShut {NoStop}%
\bibitem [{\citenamefont {Khuller}\ \emph {et~al.}(1996)\citenamefont
  {Khuller}, \citenamefont {Raghavachari},\ and\ \citenamefont
  {Young}}]{khuller1996low}%
  \BibitemOpen
  \bibfield  {author} {\bibinfo {author} {\bibfnamefont {Samir}\ \bibnamefont
  {Khuller}}, \bibinfo {author} {\bibfnamefont {Balaji}\ \bibnamefont
  {Raghavachari}}, \ and\ \bibinfo {author} {\bibfnamefont {Neal}\ \bibnamefont
  {Young}},\ }\bibfield  {title} {\enquote {\bibinfo {title} {Low-degree
  spanning trees of small weight},}\ }\href {\doibase
  10.1137/S0097539794264585} {\bibfield  {journal} {\bibinfo  {journal} {SIAM
  Journal on Computing}\ }\textbf {\bibinfo {volume} {25}},\ \bibinfo {pages}
  {355--368} (\bibinfo {year} {1996})}\BibitemShut {NoStop}%
\bibitem [{\citenamefont {Jothi}\ and\ \citenamefont
  {Raghavachari}(2009)}]{jothi2009degree}%
  \BibitemOpen
  \bibfield  {author} {\bibinfo {author} {\bibfnamefont {Raja}\ \bibnamefont
  {Jothi}}\ and\ \bibinfo {author} {\bibfnamefont {Balaji}\ \bibnamefont
  {Raghavachari}},\ }\bibfield  {title} {\enquote {\bibinfo {title}
  {Degree-bounded minimum spanning trees},}\ }\href {\doibase
  https://doi.org/10.1016/j.dam.2008.03.037} {\bibfield  {journal} {\bibinfo
  {journal} {Discrete Applied Mathematics}\ }\textbf {\bibinfo {volume}
  {157}},\ \bibinfo {pages} {960--970} (\bibinfo {year} {2009})}\BibitemShut
  {NoStop}%
\bibitem [{\citenamefont {Bui}\ and\ \citenamefont
  {Zrncic}(2006)}]{bui2006ant}%
  \BibitemOpen
  \bibfield  {author} {\bibinfo {author} {\bibfnamefont {Thang~N}\ \bibnamefont
  {Bui}}\ and\ \bibinfo {author} {\bibfnamefont {Catherine~M}\ \bibnamefont
  {Zrncic}},\ }\bibfield  {title} {\enquote {\bibinfo {title} {An ant-based
  algorithm for finding degree-constrained minimum spanning tree},}\ }in\ \href
  {\doibase 10.1145/1143997.1144000} {\emph {\bibinfo {booktitle} {Proceedings
  of the 8th Annual Conference on Genetic and Evolutionary Computation}}},\
  \bibinfo {organization} {ACM}\ (\bibinfo  {publisher} {Association for
  Computing Machinery},\ \bibinfo {address} {New York, NY, USA},\ \bibinfo
  {year} {2006})\ pp.\ \bibinfo {pages} {11--18}\BibitemShut {NoStop}%
\bibitem [{\citenamefont {Bui}\ \emph {et~al.}(2011)\citenamefont {Bui},
  \citenamefont {Deng},\ and\ \citenamefont {Zrncic}}]{bui2011improved}%
  \BibitemOpen
  \bibfield  {author} {\bibinfo {author} {\bibfnamefont {Thang~N}\ \bibnamefont
  {Bui}}, \bibinfo {author} {\bibfnamefont {Xianghua}\ \bibnamefont {Deng}}, \
  and\ \bibinfo {author} {\bibfnamefont {Catherine~M}\ \bibnamefont {Zrncic}},\
  }\bibfield  {title} {\enquote {\bibinfo {title} {An improved ant-based
  algorithm for the degree-constrained minimum spanning tree problem},}\ }\href
  {\doibase 10.1109/TEVC.2011.2125971} {\bibfield  {journal} {\bibinfo
  {journal} {IEEE Transactions on Evolutionary Computation}\ }\textbf {\bibinfo
  {volume} {16}},\ \bibinfo {pages} {266--278} (\bibinfo {year}
  {2011})}\BibitemShut {NoStop}%
\bibitem [{\citenamefont {Krishnamoorthy}\ \emph {et~al.}(2001)\citenamefont
  {Krishnamoorthy}, \citenamefont {Ernst},\ and\ \citenamefont
  {Sharaiha}}]{krishnamoorthy2001comparison}%
  \BibitemOpen
  \bibfield  {author} {\bibinfo {author} {\bibfnamefont {Mohan}\ \bibnamefont
  {Krishnamoorthy}}, \bibinfo {author} {\bibfnamefont {Andreas~T}\ \bibnamefont
  {Ernst}}, \ and\ \bibinfo {author} {\bibfnamefont {Yazid~M}\ \bibnamefont
  {Sharaiha}},\ }\bibfield  {title} {\enquote {\bibinfo {title} {Comparison of
  algorithms for the degree constrained minimum spanning tree},}\ }\href
  {\doibase 10.1023/A:1011977126230} {\bibfield  {journal} {\bibinfo  {journal}
  {Journal of heuristics}\ }\textbf {\bibinfo {volume} {7}},\ \bibinfo {pages}
  {587--611} (\bibinfo {year} {2001})}\BibitemShut {NoStop}%
\end{thebibliography}%
